\documentclass[draft,jgrga]{agutex}

 \usepackage{lineno}
 \linenumbers*[1]

\authorrunninghead{FLESHMAN ET AL.}

\titlerunninghead{SATURN'S NEUTRAL CLOUDS}

\authoraddr{F.\ Bagenal, Laboratory for Atmospheric and Space Physics,
University of Colorado
Boulder, Colorado 80309-0391}

\authoraddr{T.\ Cassidy, Laboratory for Atmospheric and Space Physics,
University of Colorado
Boulder, Colorado 80309-0391}

\authoraddr{P.\ A.\ Delamere, Laboratory for Atmospheric and Space Physics,
University of Colorado
Boulder, Colorado 80309-0391}

\authoraddr{B.\ L.\ Fleshman, 
Homer L.\ Dodge Department of Physics and Astronomy, The University of
Oklahoma, 440 W.\ Brooks St., Norman, OK 73019, USA.
(fleshman@nhn.ou.edu)}

\usepackage{color}
\usepackage{amssymb,amsmath}
\usepackage[final]{epsfig}
\graphicspath{{/Users/fleshman/Documents/research/papers/neutCloud/figs/}}
\newcommand{\Rs}{\ensuremath{\textrm{R}_\textrm{S}}}
\newcommand{\kms}{\ensuremath{\textrm{km\ s}^{-1}}}
\newcommand{\ms}{\ensuremath{\textrm{m\ s}^{-1}}}
\newcommand{\kgs}{\ensuremath{\textrm{kg\ s}^{-1}}}
\newcommand{\extrap}{\ensuremath{\sigma^\textrm{extrapolated}_\textrm{OH}}}
\newcommand{\sym}{\ensuremath{\sigma^\textrm{symmetric}_\textrm{OH}}}
\newcommand{\asym}{\ensuremath{\sigma^\textrm{asymmetric}_\textrm{OH}}}
\newcommand{\icarus}{Icarus}
\newcommand{\nat}{Nature}
\newcommand{\jcp}{Journal of Chemical Physics}
\newcommand{\planss}{Planetary Space Science}
\newcommand{   \solphys     }{ Solar Physics }
\begin{document}

%
%

\title{The roles of charge exchange and dissociation in spreading Saturn's neutral clouds}

%

%
%



\authors{B.\ L.\ Fleshman, P.\ A.\ Delamere, F.\ Bagenal, T.\ Cassidy}







%
%


\begin{abstract}
Neutrals sourced directly from Enceladus's plumes are initially confined to a dense
neutral torus in Enceladus's orbit around Saturn.  This neutral torus
is redistributed by charge exchange, impact/photodissociation,
and neutral--neutral collisions to produce Saturn's neutral clouds. 
Here we consider the former processes in greater detail than in previous
studies.  In the case of dissociation, models have assumed that OH is produced
with a single speed of 1 \kms, whereas laboratory measurements suggest a range
of speeds between 1 and 1.6 \kms.  We show that the high-speed case increases
dissociation's range of influence from 9 to 15 \Rs.  For charge exchange,
we present a new modeling approach, where the ions are followed within a neutral
background, whereas neutral cloud models are conventionally constructed from the
neutrals' point of view.  This approach allows us to comment on the
significance of the ions' gyrophase at the moment charge exchange occurs.
Accounting for gyrophase: (1) has no consequence on the H$_2$O cloud; (2) doubles
the local density of OH at the orbit of Enceladus; and (3) decreases the
oxygen densities at Enceladus's orbit by less than 10\%.
Finally, we
consider velocity-dependent, as well as species-dependent cross sections
and find that the oxygen cloud produced from charge exchange is spread out more 
than H$_2$O, whereas the OH cloud is the most confined.

\end{abstract}

%
%

%

\begin{article}

%
%
%
\section{Introduction}
The Enceladus plumes directly produce a dense
H$_2$O torus centered on Enceladus's orbit, within which
charge exchange and dissociation subsequently produce neutrals that either
feed Saturn's extended neutral clouds, collide (absorb) with Saturn and its
rings, or leave the system altogether on escape orbits.  
This paper is a report on the results of a sensitivity study of low-velocity
charge exchange and dissociation within the neutral torus.

Several decades before Cassini arrived at Saturn and the Enceladus water plumes were
discovered \citep{hansen2006},
neutral hydrogen was observed in Saturn's magnetosphere,
both from Earth \citep{weiser1977} and from Voyagers 1 and 2
\citep{shemansky1992}.  Hydroxyl was later discovered by \cite{shemansky1993} 
with HST, and more recently, \cite{esposito2005} detected atomic oxygen.
These observations collectively hinted
at the presence of a source of water, and models predicted its location to be
near the orbit of Enceladus (c.f., \cite{jurac2002}).

After identifying the Enceladus plumes as the dominant source of the
water-group neutrals (O, OH, H$_2$O)---and indeed the co-rotating plasma itself via
electron impact and photoionization
\citep{young2005,sittler2005,sittler2008}---researchers have been attempting to understand how
neutrals are transported from Enceladus to 20 Saturn radii (\Rs\ = 6 $\times10^9$ 
cm) and beyond, as observed by
\cite{shemansky1993}, \cite{esposito2005}, and most
recently by \cite{melin2009}.  Early on,
\cite{jurac2002} mentioned the role of charge exchange in
this inflation process.  \cite{johnson2006} later showed that if
magnetospheric plasma is slowed sufficiently with respect to
neutrals in the Enceladus torus,
then charge exchange produces a sufficient number of particles with velocities
capable of spreading the dense H$_2$O Enceladus torus into the cloud 
observed by \cite{shemansky1993}.

\cite{farmer2009a} pointed out the importance of dipole--dipole
interactions in collisions involving H$_2$O molecules.  She
showed that collisions inside the dense
Enceladus torus (parameterized by macroscopic viscosity) are alone sufficient 
for the creation of the extended component of Saturn's neutral
cloud.  \cite{cassidy2010a} later argued that Farmer's fluid treatment
is inappropriate for neutral--neutral collisions in the Enceladus torus,
where the mean free path is on the order of the torus size itself.
Instead, \cite{cassidy2010a} approached the problem
with a direct simulation Monte Carlo (DSMC) model. Their model self-consistently
included losses due to charge exchange, dissociation, and ionization, whereas
\cite{farmer2009a} accounted for losses to charge exchange and ionization by
evolving the neutral cloud for the time scales (months to a few years) found in \cite{sittler2008}.
Both studies agree that
neutral--neutral collisions are necessary for the inflation of Saturn's
neutral cloud.

Collisions between neutrals occur at a rate proportional to the square of
the neutral density.  Thus, where neutral densities peak near the orbit of Enceladus, 
neutral--neutral collisions occur more often than either charge exchange
or dissociation,
whereas near 6 \Rs, neutral densities drop and all three processes
become comparable (see Fig.\ 3, \cite{cassidy2010a}).  Models involving neutral
collisions have recently been validated with Herschel observations by
\cite{hartogh2011}, who attribute a warm and broadened Enceladus torus to
heating via neutral--neutral collisions;
the effect of these interactions should therefore be included in 
any attempt to fully model Saturn's neutral clouds.
Nevertheless, several first-order conclusions can be drawn by
revisiting charge exchange and dissociation.

Previous neutral cloud models approach charge exchange from the neutrals' point of view, whereas
we follow the ion along its trajectory, thus allowing 
us to identify the gyrophase at which an ion undergoes charge exchange.
We find that including the phase
dependence doubles OH densities at the orbit of Enceladus, decreases oxygen
density by $\lesssim$ 10\%, and has no effect on H$_2$O (section
\ref{chargeExchangeResults}).
Also, the velocity-dependence of charge exchange varies by species.
Previous studies ($e.g.$,
\cite{johnson2006,cassidy2010a}) have considered velocity-dependence, but have used a single
cross section to represent all charge exchanges.
We show in section  \ref{chargeExchangeResults} that symmetric reactions 
such as $\mathrm{H_2O+H_2O^+\rightarrow H_2O^++H_2O^*}$
(the asterisk identifies a neutral released with the speed of the reacting ion) 
tend to distribute neutrals closer to
Saturn,
while
asymmetric exchanges such as $\mathrm{H_2O+O^+\rightarrow H_2O^++O^*}$
populate a more extended cloud, with less 
absorption on Saturn.

With regard to dissociation, OH produced by impact/photodissociation of H$_2$O
has previously 
been modeled with an initial speed of 1 \kms\ \citep{jurac2005,cassidy2010a}, 
whereas recent laboratory measurements
span 1 to 1.6 \kms,
depending on the molecule's internal energy \citep{wu1993,makarov2004}.  
We model this parameter space
and find that, relative to charge exchange, most OH found inside  9 \Rs\ is
produced by dissociation when OH is dissociated from H$_2$O at 1 \kms, with that location extended to 15 \Rs\ when OH is
dissociated from H$_2$O at 1.6 \kms\ instead.

This paper is organized as follows.  The model for the production of neutrals
via dissociation and three illustrative charge exchanges is explained in
section \ref{model}.  Our results
are found in section \ref{results}, followed by a discussion in section
\ref{discussion}.  The important points are summarized in section \ref{conclusions}.

%
\section{Model}\label{model}
We begin with a few words on nomenclature.  The neutral torus in this paper
pertains to the
primary neutral torus (not plasma torus) supplied directly by Enceladus's
plumes.  The neutral clouds refer to the secondary neutrals produced from charge
exchange and dissociation in the neutral torus.

The production of Saturn's neutral cloud is modeled
in two steps.  We first construct a dense H$_2$O torus from a 
plume positioned at Enceladus's south pole with
specifications based on several Cassini Enceladus flybys (\cite{smith2010}; see
also \cite{smith2004}, \cite{smith2006}, and \cite{smith2007}).  Secondary
neutrals are then produced from the primary neutral torus by charge exchange and dissociation,
some of which remain gravitationally bound to Saturn and form the neutral clouds.  On the
basis that they spend most of the time outside the neutral torus and plasma
sheet, 
we assume their lifetimes to be determined solely by photo-processes,
though in section \ref{results:neutralFates}, we consider the effects of
charge exchange and electron impact.
\subsection{Neutral torus model}\label{neutralTorusProduction}
The neutral torus and Enceladus plume models are described in the following
subsections.
%
\subsubsection{Enceladus H$_2$O torus} 
\label{enceladusTorus}
Our aim is to study the effects of several important reactions occurring in Enceladus's orbit.
The primary neutral torus, fed directly by Enceladus, is produced in the model by
evolving
water molecules released from Enceladus into a dense neutral torus
centered on Enceladus's orbit (3.95\,\Rs).
The assumption is that all H$_2$O is initially produced by a single plume at
Enceladus's south pole.  In reality, more than one plume has been
observed \citep{porco2006}, and researchers such as \cite{saur2008} and \cite{smith2010} 
have studied their signature on flyby observations.
For our purposes, the detailed influence of multiple plumes can be neglected.
The plume particles' radial speed distribution is prescribed as a
one-dimensional
Maxwellian with temperature $T=180$\,K \citep{spencer2006,hansen2006}:  
\begin{linenomath}
\begin{equation}
f(v)=\left(\frac{m_\mathrm{H_2O}}{2\pi kT}\right)^{1/2} 
\exp\left[-\frac{m_\mathrm{H_2O}}{2kT}(v-v_\mathrm{bulk})^2\right],
\label{speedDist}
\end{equation}
\end{linenomath}
where $v_\mathrm{bulk}$ is the
bulk speed, equal to 720\,m\,s$^{-1}$, 1.8$\times$ the thermal speed 
estimated by \cite{smith2010} ($v_\mathrm{therm}=\sqrt{2kT/m_\mathrm{H_2O}}=400$\,\ms).
Additionally, a raised cosine distribution is
used to determine where the molecules are released:
\begin{linenomath}
\begin{equation}
g(\theta)=\left\{ \begin{array}{ll}
\frac{1}{\theta_0}\left[1+\cos\left(\frac{\theta}{\theta_0}\pi\right)\right]
 &\mbox{ if $\theta < \theta_0=30^\circ$} \\
0 &\mbox{ otherwise.}
       \end{array} \right. 
\label{cosineDist}
\end{equation}
\end{linenomath}
Co-latitude theta is measured from Enceladus's south pole, and 
$\theta_0=30^\circ$ is the plume half-width,
based on INMS in situ observations \citep{smith2010}.  We
assume no azimuthal dependence.

Enceladus's gravity is ignored since the escape
velocity, $v_\mathrm{esc}\,=\,
240 \mathrm{\ m\ s^{-1}}$, is greatly exceeded for most molecules (99\%), where
$v_\mathrm{bulk}>v_\mathrm{therm}>v_\mathrm{esc}$;
our results differ by less than one percent
whether Enceladus's gravity is considered or not.  
Particles released from Enceladus are thus assumed to move on 
Keplerian orbits with respect to Saturn. 
Each water molecule is allowed to orbit 
inside the torus for a period determined by the collective lifetimes against
photodissociation, electron-impact dissociation, and charge exchange.
To be clear, the molecules forming the neutral torus are subject 
to all of the losses stated,
while the neutral cloud is subject to 
photo-processes only (section \ref{model:neutralCloudProduction}). 
Further reaction details are important for modeling plasma characteristics 
but only the timescales given below are required to model the neutral torus.

\paragraph{Photodissociation}  The photodissociation lifetime 
for H$_2$O, $\tau_\mathrm{phot}=9.1\times 10^6$ s, 
comes directly from \cite{huebner1979}, scaled to Saturn's distance from the
Sun.  At peak solar activity, 
neutral abundances attributed to dissociation double
\citep{jackman2011}. 

\paragraph{Impact dissociation}  In the case of impact dissociation,
suprathermal (hot) electrons
dominate \citep{fleshman2010b}.  Assuming the conditions near Enceladus's orbit
apply throughout the neutral torus,
we estimate the hot electron density and temperature to be 160 eV and 0.3 cm$^{-3}$
from \cite{fleshman2010b}, which fit within a range of recent
observations (cf., \cite{young2005}, \cite{sittler2008}).
The rate coefficient for impact dissocation of water is found by convolving
$\sigma(v)v$ with a 160 eV Maxwellian distribution
to find $\kappa_\mathrm{imp}=1.5\times10^{-6}\mbox{ cm$^3$ s$^{-1}$}$
(\cite{fleshman2010b}, table S9).
Above 100 eV, $\kappa_\mathrm{imp} $ is insensitive to 
temperature, making this estimate valid for a range of observations.
Assuming the hot electron density,
$n_\mathrm{eh}$, is constant over the neutral torus, the lifetime
against impact dissociation is 
$\tau_\mathrm{imp}\approx[\kappa_\mathrm{imp}n_\mathrm{eh}]^{-1}=2.2 \times 10^6$ s.
Dissociation $via$ thermal electrons---whose
temperature and density are 2 eV and 60 cm$^{-3}$---is also expected, but such
collisions occur 5$\times$ less often (\cite{fleshman2010b}, table S9) 
and are thus ignored here.

\paragraph{Charge exchange} The following three reactions are included in this
study:
\begin{linenomath}
\begin{subequations}
\begin{align}
\mathrm{H_2O+H_2O^+} &\rightarrow \mathrm{H_2O^+ +H_2O^*} \label{h2o_chex}\\
\mathrm{H_2O+H_2O^+} &\rightarrow \mathrm{H_3O^+ +OH^*} \label{oh_chex}\\
\mathrm{H_2O+O^+} &\rightarrow \mathrm{H_2O^++O^*} \label{o_chex}.
\end{align}
\end{subequations}
\end{linenomath}
Other charge exchanges are important in the neutral torus, some of which
involve ions reacting with secondary neutrals such as H, O, and OH.  
To model their effect properly, one would calculate these neutral densities as
in a conventional time-dependent neutral cloud model.  We estimate that including all such
reactions would increase our estimates on neutral cloud densities by
approximately a factor
of two.  The primary purpose of choosing this combination of reactions is to
study three classes of charge exchanges, for which the collision
frequency decreases with, increases with, or is independent of the 
relative speed of the reacting pair (reactions \ref{oh_chex}, \ref{o_chex}, and \ref{h2o_chex},
respectively).  We return to this point in section \ref{chargeExchange}.

An estimate of the charge exchange 
lifetime can be made by adding the rate
coefficients for reactions \ref{h2o_chex}--\ref{o_chex}.
Multiplying by the observed H$_2$O$^+$ and O$^+$ densities near the orbit of Enceladus
 (6 and 12 cm$^{-3}$, \cite{sittler2008}), we find 
$\tau_\mathrm{chex}=[\kappa_\mathrm{exch}^\mathrm{H_2O^+}n_\mathrm{H_2O^+}
                    +\kappa_\mathrm{exch}^\mathrm{O^+}n_\mathrm{O^+}]^{-1}
      =[6.0+2.8]^{-1}\times10^8 \mathrm{\ s\ }=1.1\times 10^7$ s.  
The reaction rates are from \cite{lishawa1990} and \cite{albritton1978} for
$\kappa^\mathrm{H_2O^+}_\mathrm{exch}$ and $\kappa^\mathrm{O^+}_\mathrm{exch}$,
respectively.
The lifetime of H$_2$O against the sum of these processes is then 
\begin{linenomath}
\begin{eqnarray} 
\tau_\mathrm{torus}=
\left[\frac{1}{\tau_\mathrm{phot}} +
      \frac{1}{\tau_\mathrm{imp}} +
      \frac{1}{\tau_\mathrm{chex}}\right]^{-1} &=&
(1.1+4.5+0.91)^{-1}\times10^7 \mathrm{\ s} \nonumber \\ 
&=& 1.6\times10^6 \mathrm{\ s}
\approx 20 \mbox{ days.} \label{t_total}
\end{eqnarray}
\end{linenomath}
These estimates can be compared to Fig.\ 3 of
\cite{cassidy2010a}:  for example, our lifetime against dissociation is 1.7
$\times 10^6$ s, while they use  $7\times10^6$ s near 4\,\Rs.
The discrepancy comes mostly from the impact dissociation timescale.
Cassidy's dissociation rate was calculated using \cite{Schippers2008},
wherein CAPS ELS data were fitted and extapolated down from 5.5 \Rs, while our
own estimate hinges on a hot electron density derived from our chemistry model
\citep{fleshman2010b}.
For charge exchange, we have a lifetime of 1.1 $\times10^7$ s, and
\cite{cassidy2010a} have a comparable $8\times10^6$ s.

Particles are created and tracked in each of our
model runs, and the results are
scaled to the number of water molecules in the real neutral torus.
The total number is estimated from an assumed
neutral source rate from Enceladus of
$\dot{M}=200\ \kgs$ \citep{jurac2005,hansen2006,hansen2011}
 and lifetime, $\tau_\mathrm{torus}$ (Eq.\ \ref{t_total}):
\begin{linenomath}
\begin{equation}
N_\mathrm{torus}=
\dot{M}\tau_\mathrm{torus}/m_\mathrm{H_{2}O}\approx 1.1\times 10^{34} \mathrm{\ H_2O\ molecules}.
\label{torusNeutrals}
\end{equation} 
\end{linenomath}
We bin and azimuthally average the results
to find a 2-D density function, $n_\mathrm{torus}(r,\theta)$ (radius and
latitude), through which to
introduce ions for charge exchange.  This function also
determines from where dissociated neutrals are produced.

\subsubsection{Plume model} 
\label{plume}
Before describing dissociation and charge exchange within the torus, we address a
calculation with the purpose of comparing the neutral production
near Enceladus with that from the entire
torus.  In doing so, we prescribe a plume whose density is
consistent with Eqs.\ \ref{speedDist} and \ref{cosineDist}.
In this case, the ambient neutral density can be ignored compared to the neutrals leaving
the surface of Enceladus directly.

The densities are determined everywhere by imposing integrated flux
($\int n(r,\theta)v(r,\theta) dA$) and energy ($mv^2/2-mM_\mathrm{E}G/r$)
conservation at a given distance $r$.  The picture can be simplified, however,
since most neutrals have at least twice Enceladus's escape velocity.
The speeds are thus independent of $r$ in the immediate vicinity of Enceladus.
By equating the integrated flux at
the surface of Enceladus to the same integral at another distance
$r>\mathrm{R_E}$, we find
a familiar $1/r^2$ dependence:
\begin{linenomath}
\begin{equation}
\label{plume_density}
 n_\mathrm{plume}(r,\theta)=n(\theta)\left(\frac{\mathrm{R_E}}{r}\right)^2\exp\left[
-\left(\frac{r-\mathrm{R_E}}{H_r}\right)\right].
\end{equation}
\end{linenomath}
The trailing exponential factor is imposed $ad$ $hoc$ to keep the total plume content finite,
and reflects Saturn's influence as the molecules leave Enceladus.
Consistent with \cite{saur2008}, $H_r$ is set at 4 times the Hill radius of 948\,km.
The angular dependence is consistent with our velocity
distribution in the previous section (Eq.\ \ref{speedDist}), 
\begin{linenomath}
\begin{equation}
n(\theta)=\left\{ \begin{array}{ll}
\frac{n_0}{2}\left[1+\cos\left(\frac{\theta}{\theta_0}\pi\right)\right]
 &\mbox{ if $\theta < \theta_0=30^\circ$} \\
0 &\mbox{ otherwise,}
\end{array} \right.
\end{equation}
\end{linenomath}
which is normalized such that
$n(0)=n_0$, where the plume strength $n_0$ is found to be
$5.9\times10^8\ \mathrm{cm}^{-3}$ by integrating
$n(\theta)v_\mathrm{bulk}$ over the area spanning the south pole of Enceladus from
$\theta=0^\circ$ to $\theta=\theta_0=30^\circ$, and setting that result equal to the
plume production rate of $\dot{M}/m_\mathrm{H_2O}=(200\ \kgs)/m_\mathrm{H_2O}=6.7\times10^{27}$
molecules per second.

In section \ref{results}, we compare the results of charge exchange with the plume 
($n_\mathrm{plume}$) to that with the entire torus ($n_\mathrm{torus}$).

\subsection{Neutral cloud model} \label{model:neutralCloudProduction}
Production of Saturn's neutral clouds entails following the neutrals produced by
dissociation and charge exchange occurring within the neutral torus.  The
treatment of each of these
processes are described below.

\subsubsection{Dissociation} \label{dissociation}
The hydroxyl radical, OH, produced largely by
dissociated H$_2$O, has previously been 
modeled with a single speed of 1 \kms\
(\textit{i.e.}, \cite{jurac2005}, \cite{cassidy2010a}).
Dissociated OH has been measured however with speeds between 1
and 1.6 \kms\ \citep{wu1993,makarov2004}. 
Here we bound this range by modeling the OH neutral clouds produced from
an azimuthally-symmetric source (with respect to Saturn) with velocities drawn from Maxwellian
distributions with temperatures
$T=\frac{1}{2}m_\mathrm{OH}v_\mathrm{mp}^2$, where the most probable speed,
$v_\mathrm{mp}$, is set to 1 and
1.6 \kms, representing the low- and high-speed limits.

The initial locations of the ejected OH are determined by the spatial distribution of
neutrals in the Enceladus torus ($n_\mathrm{torus}(r,\theta)$,
section \ref{enceladusTorus}), and the
directions of their release are chosen randomly and isotropically.  The molecules orbit
Saturn until they are photodissociated and removed from
the system.

By assuming a volume over which dissociations occur, 
the number of modeled
OH molecules can be scaled to a realistic value.  We take the volume to be a torus
centered on Enceladus (3.95 \Rs), with a minor
radius of 1 \Rs:
\begin{linenomath}
\begin{equation} \label{volume}
V\approx 2\pi(4\,\Rs)(2\,\Rs)^2=2\times10^{31}\,\mathrm{cm}^3.
\end{equation}
\end{linenomath}
For impact dissociation, we then expect a contribution of 
\begin{linenomath}
\begin{equation} \label{impDissOH}
N^\mathrm{imp}_\mathrm{cloud}=k_\mathrm{imp}\tau_\mathrm{phot}^\mathrm{OH}V
=2.8\times 10^{34} \mbox{\ OH molecules},
\end{equation}
\end{linenomath}
where $k_\mathrm{imp}=7.9\times10^{-5}$ cm$^{-3}$ s$^{-1}$ 
is the rate (per volume) of impact dissociations occurring between suprathermal electrons and H$_2$O
molecules in the torus \citep{fleshman2010b} and  
$\tau_\mathrm{phot}^\mathrm{OH}=1.8\times 10^7$ s is the photodissociation
lifetime of OH at Saturn \citep{huebner1979}.
The number of OH molecules produced by photodissociation in the
torus is similarly given by
\begin{linenomath}
\begin{equation} \label{photDissOH}
N^\mathrm{phot}_\mathrm{cloud}=k_\mathrm{phot}\tau_\mathrm{phot}^\mathrm{OH}V
=7.6\times 10^{33} \mbox{\ OH molecules},
\end{equation}
\end{linenomath}
where $k_\mathrm{phot}=2.1\times10^{-5}$ cm$^{-3}$ s$^{-1}$ 
is the rate (per volume) of H$_2$O photodissociations 
occurring in the Enceladus torus (\cite{fleshman2010b}, Table S9).  
The total abundance attributed to dissociation is then given by the sum of Eqs.\
\ref{impDissOH} and \ref{photDissOH}.
\cite{cassidy2010a} constrained their study with HST observations
\citep{melin2009} and found a similar OH content (see comparison in Fig.\
\ref{fleshmanVcassidy}c, this paper).

That neutral production by photo- and impact dissociation are comparable 
in magnitude is itself noteworthy.  This condition is 
not shared by systems with hotter and denser plasma.  For example,
electron impact dissociation and ionization dominate over photon-driven processes
in Jupiter's Io torus, where the plasma is warmer where the pick-up energies are four times
higher than at Enceladus
\citep{delamere2007,fleshman2010b}.  We also note that unlike with Io,
long neutral lifetimes
in the Enceladus neutral torus inhibit the response of Saturn's neutral clouds
to short-term plume variability, though variability on the order of months has
been studied by \cite{smith2010}.

\subsubsection{Charge exchange} \label{chargeExchange}
We now describe the model for producing and following neutrals from charge
exchange.
\cite{cassidy2010a} and \cite{jurac2005} also considered velocity-dependent
charge exchange, but unlike these previous studies, we capture the gyrophase
at which the reactions occur by following ions along their trajectories (section
\ref{chargeExchangeResults}).
We also prescribe cross sections specific to each reaction,
being particularly interested in the effects of low-velocity charge exchange.

At very high speeds, the cross sections go to zero for all charge
exchanges \citep{johnson1990}.  At low relative velocities, however (few
\kms), the details of the collision are determined by the nature of the reacting species.  If
the reactants and products are identical, apart from an electron ($i.e.$, 
$\mathrm{H_2O + H_2O^+}\rightarrow \mathrm{H_2O^+ + H_2O^*}$), 
the reaction is termed resonant, or symmetric, and the cross
sections grow as the inverse of the relative speed.  
If the reactants differ, as with 
$\mathrm{H_2O + O^+}\rightarrow \mathrm{H_2O^+ + O^*}$,
the cross sections are likely to vanish at low speeds
\citep{rapp1962}---the difference being that the energy of the electron configurations
is unchanged for symmetric-type charge exchanges
\citep{johnson1990}.  
Neutrals produced from resonant charge exchange therefore tend to have lower
velocities than do neutrals produced from
non-resonant (asymmetric) charge exchange.  This is a key point central to much of
our discussion in section \ref{results}.

\begin{figure}
 \noindent\includegraphics[width=6.5in]{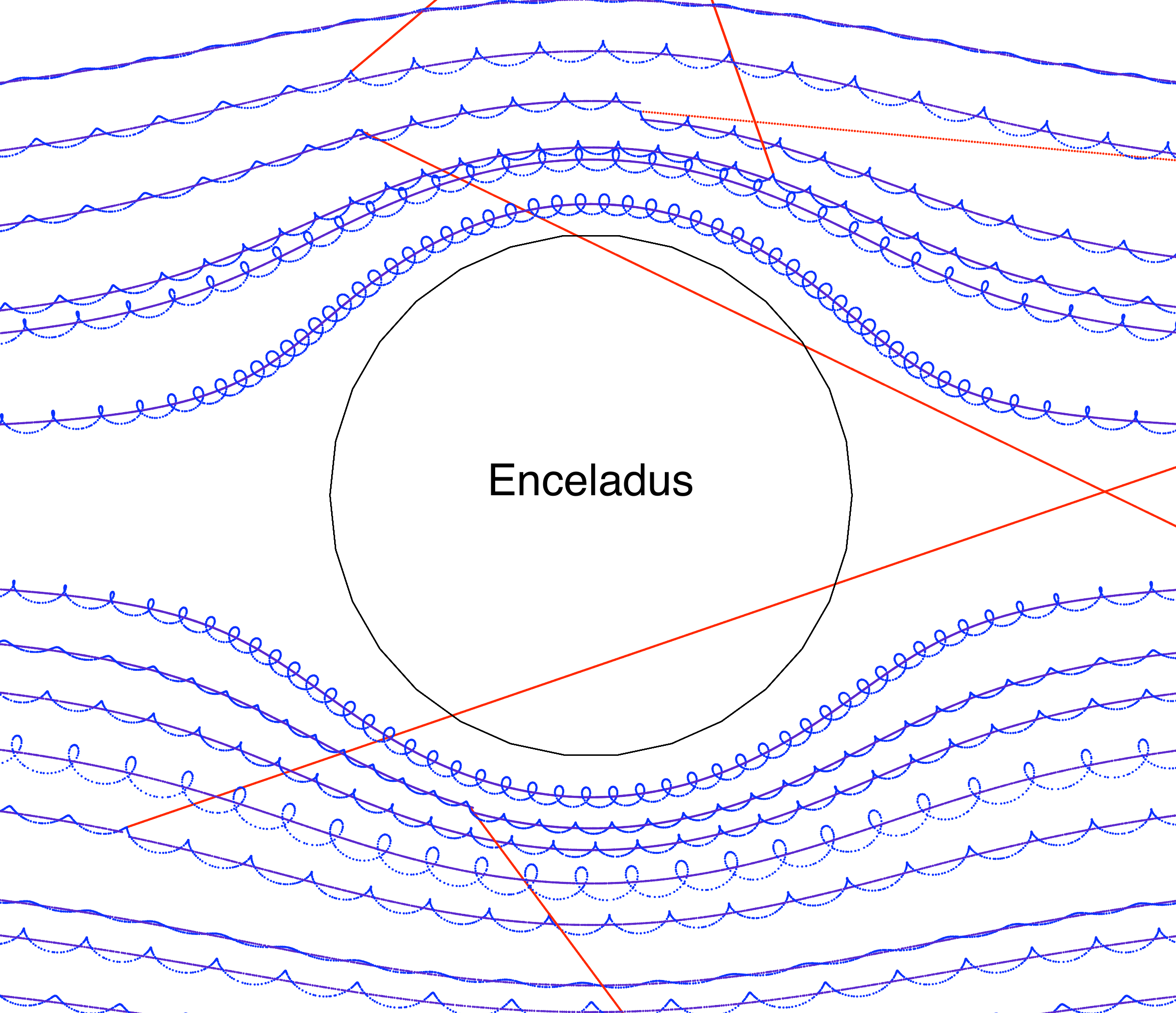}
 \caption{Sketch of gyrating ions in the neutral frame
with guiding centers moving along a prescribed
flow field, shown here near Enceladus for scale.
Warm ions ($v_\perp > v_\mathrm{flow}$) move on trajectories that coil 
around themselves and do not reach zero relative velocity with respect to the
neutrals at any point.
Cool ions ($v_\perp < v_\mathrm{flow}$) essentially trace
their guiding centers with `snake-like' trajectories, 
and also do not obtain zero relative velocity.  Fresh pick-up ions 
($v_\perp \approx v_\mathrm{flow}$) do, however, obtain zero relative velocity at the
cusps of their cycloidal trajectories.  Neutrals produced by charge exchange
(whose trajectories are indicated by the red lines) tend to be created
with velocities at which the respective reaction rates peak (Fig.\
\ref{reacRates}).}
 \label{moleculesFromChex}
 \end{figure}

Individual ions are followed as they traverse the neutral torus
(section \ref{neutralTorusProduction}).  This approach allows their gyrophase
to be determined the instant that
charge exchanges occur (see Fig.\ \ref{moleculesFromChex}).  
The implicit assumption is that the collision is elastic, and that the neutral
product has an initial velocity given by the ion velocity just before the 
exchange takes place.

The ions are introduced into the model from two Maxwellian speed distributions,
\begin{linenomath}
\begin{align}
 f_\perp(v_\perp)&=
  \frac{m_\mathrm{ion}}{kT_\perp}v_\perp 
  \exp\left[-\frac{m_\mathrm{ion}v_\perp^2}{2kT_\perp} \right]
  \mbox{ (speeds perpendicular to $B$)}\\
 f_\parallel(v_\parallel)&=
  \sqrt{\frac{m_\mathrm{ion}}{2\pi kT_\parallel}}
  \exp\left[-\frac{m_\mathrm{ion}v_\parallel^2}{2kT_\parallel} \right]
  \mbox{ (speeds parallel to $B$)}
\end{align}
\end{linenomath}
with a temperature anisotropy of
\begin{linenomath}
\begin{equation}
\frac{kT_\perp}{kT_\parallel}
=\frac{27\,\mathrm{eV}}{5.4\,\mathrm{eV}}=5
\end{equation}
\end{linenomath}
for both O$^+$ and H$_2$O$^+$ \citep{sittler2008}.
The perpendicular temperature is derived from the pick-up ion velocity
at the orbit of Enceladus, determined from CAPS data by \cite{wilson2009}
[$kT_\perp=\frac{1}{2}m_\mathrm{W^+}(v_\phi-v_\mathrm{Kep})^2$].
The ions also rotate around a guiding center (field line)
moving at $v_\phi=18$ \kms\ in a frame rotating with the neutrals.

For the component of our study aimed at estimating local neutral production
(section \ref{plume}),
ions passing near Enceladus are diverted (treating
Enceladus as a rigid cylinder) and are slowed to 10\% of the ambient flow speed to
account for the effects of mass-loading (see \cite{fleshman2010a}).
 
Time steps are taken at less than 1\% of an ion's gyroperiod:
\begin{linenomath}
\begin{equation}
\Delta t=R\,\times\,T_\mathrm{gyro}=R\,\times\,\frac{2\pi m}{qB},
\label{dt}
\end{equation}
\end{linenomath}
where $R$ is a random number between 0 and 0.01, 
$T_\mathrm{gyro}$ is the ion's gyroperiod (3.6\,s for H$_2$O$^+$), $B$\,=\,325\,nT, and $q$ and
$m$ are the charge and mass, respectively, of the reacting ion.
Such resolution is necessary in order to 
capture the significance of the energy
dependence at low relative
speeds.
After each time step, the collision frequency
$\nu$ is calculated from 
\begin{linenomath}
\begin{equation}
\nu(r,\theta,v_\mathrm{rel})=n(r,\theta) \sigma(v_\mathrm{rel})v_\mathrm{rel},
\end{equation}
\end{linenomath}
where $n(r,\theta)$ is the local H$_2$O density (section
\ref{neutralTorusProduction}), 
$v_\mathrm{rel}$ is the relative velocity between the reacting ion and
neutral, and $\sigma(v_\mathrm{rel})$ is the velocity-dependent cross section.
Poisson statistics are used to test the likelihood of one or more
reactions having occurred within $\Delta t$.  If
$\exp(-\nu \Delta t)$ is less than a second random number between 0 and 1,
then a reaction occurs.  The possibility of multiple
reactions occurring over $\Delta t$ is taken into account, but it 
is neglectable (appendix A).
 
As with OH produced by dissociation (section \ref{dissociation}), neutrals produced by charge
exchange are followed under the influence of Saturn's gravity
until they are photodissociated or photoionized.
Their initial location and velocity are taken to be that of
the reacting ion, pre-transfer.

The model runs are centered on the orbit of Enceladus spanning  10 R$_\mathrm{E}$
in the direction of corotation 
(R$_\mathrm{E}$ = 250 km = radius of Enceladus) and
$\pm$ 120\,R$_\mathrm{E}$ (0.5\,\Rs) in
both the radial and $z$ directions to adequately sample the H$_2$O torus
(section \ref{enceladusTorus}).  
Ions are introduced into the model on the
upstream boundary, and their guiding centers flow downstream at a speed 
$v_\mathrm{plasma}=18$ \kms\ relative to the neutrals.
Their starting location in ($r,z$) is chosen randomly.

\paragraph{Scaling} 
The neutral clouds formed $via$ charge exchange are done so in our model 
by following a relatively small number of ions, and must thus be scaled 
to facilitate comparison with observations and other models.
The number of neutrals in our modeled clouds have
been scaled by accounting for the following.  First,
the number of representative ions used to produce the neutral clouds $via$ charge
exchange falls short of, and must be scaled to,
the number of ions present in the actual plasma torus, $n_\mathrm{ion}V$.
The volume of the plasma torus, $V$, is given in Eq.\ \ref{volume}, and
$n=12$ and 6 cm$^{-3}$ for O$^+$ and H$_2$O$^+$, respectively \citep{sittler2008}.
Second, we have argued that photo-processes
are more likely to occur than either charge exchange or
electron-impact processes throughout the neutral clouds
with the exception of very near
the neutral torus.
In keeping with this assumption, the plasma torus thus feeds the extended neutral
clouds $via$ charge exchange for a photodissociation (photoionization in the
case of oxygen) time scale before equilibrium of the neutral cloud is achieved:
$\tau_\mathrm{phot}= $ 14, 0.6, 0.3 years for O, OH, and H$_2$O, respectively
\citep{fleshman2010b}.
Our model runs followed 10$^5$ ions for 100 seconds, and 
the resulting neutral clouds were scaled as described.

\section{Results} \label{results}

In the following sections, we present and discuss the neutral clouds resulting from
dissociation and charge exchange in our model.
%
\subsection{Charge exchange} \label{chargeExchangeResults}
In the neutrals' reference frame,
ions oscillate between $\approx$ 0 \kms\  and twice the local
pick-up speed ($v_\phi\approx 18\ \kms$) due to gyro-motion.
A cartoon of this can be seen in Fig.\ \ref{moleculesFromChex},
where $v_\mathrm{rel}\approx0$ at the cusp of the ion trajectory and reaches a
maximum of $v_\mathrm{rel}\approx 2 v_\phi$ along the flow direction.  Shown are
several trajectories for which $v_\perp$ is either less than, greater than, or
approximately equal to the bulk flow velocity.  The neutrals formed $via$ charge
exchange follow the trajectories indicated in red.

The velocity dependence of reactions
\ref{h2o_chex}--\ref{o_chex} are
determined by the details of the reacting species \citep{johnson1990}.
Essentially, symmetric charge exchanges have cross sections that
increase monotonically with decreasing velocity, whereas cross sections for 
asymmetric exchange peak and then vanish at low relative speeds.  The
implication is that symmetric exchanges produce lower velocity neutrals and a
more compact neutral cloud than do asymmetric reactions.

With symmetric charge exchange, the cross sections go as
${v_\mathrm{rel}}^{-1}$, so that the collision frequency ($n\sigma v$) is
independent of $v$, as with reaction \ref{h2o_chex}, whereas
asymmetric exchanges are defined by cross sections (and collision
frequencies) which 
tend rapidly toward zero at low relative velocities ($\sim {v_\mathrm{rel}}^4$,
 \cite{rapp1962}).
 
\begin{figure}
 \noindent\includegraphics[height=6.5in]{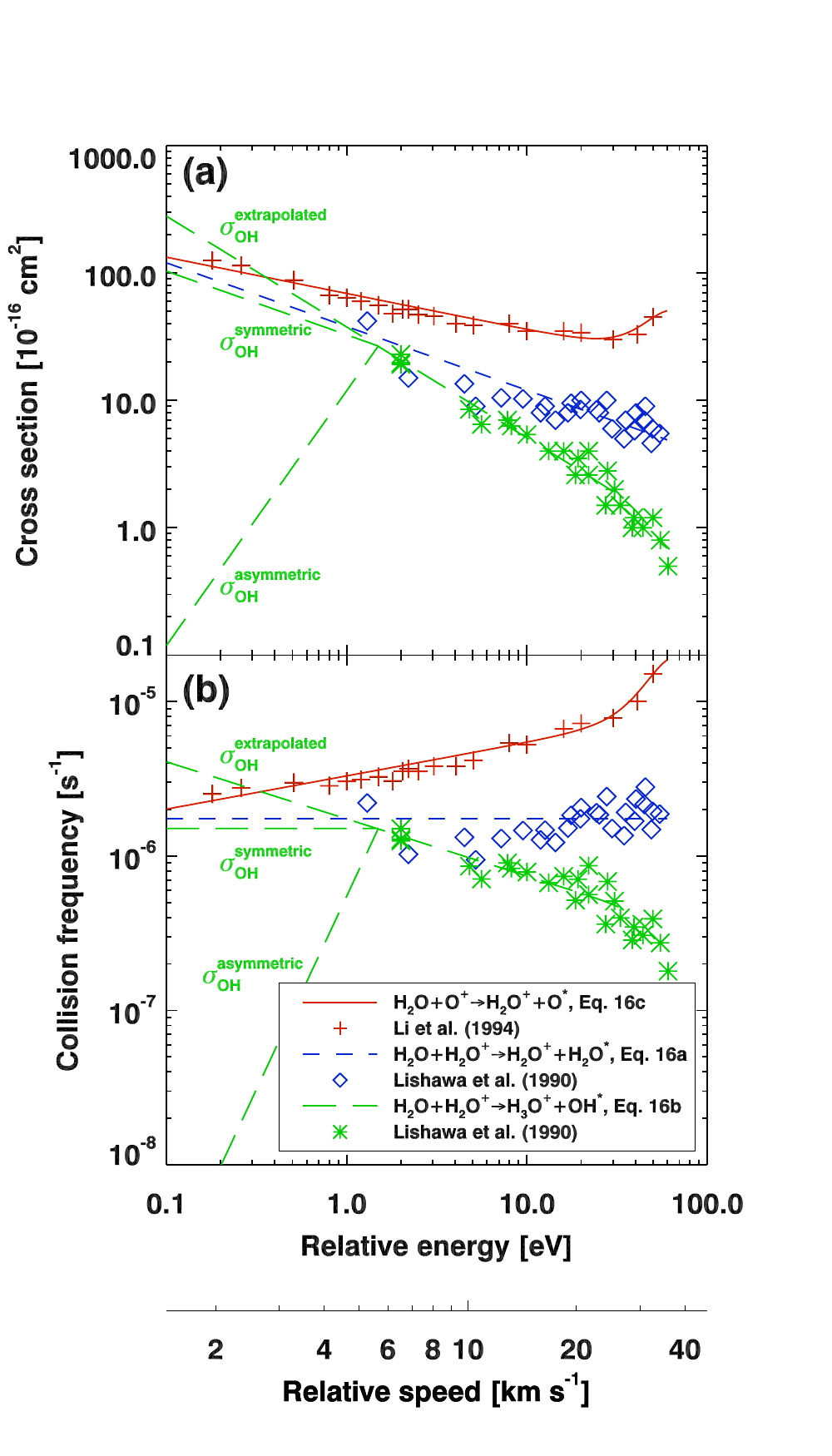}
 \caption{(a.)  Cross sections for the reactions listed in the legend.
Data for reactions \ref{h2o_chex} and \ref{oh_chex} are from \cite{lishawa1990}, and
data for reaction \ref{o_chex} is from \cite{li1994}.
\extrap, \sym, and \asym\ are hypothetical fits applying to the OH$^*$-producing reaction, and are explored in
Fig.\ \ref{lowVelChexCloud}.  
Ions oscillate between $\approx0$ and 36 \kms in the
Enceladus torus.
(b.)  Collision frequency, $n\sigma(v)v$, for a given density of
$n_\mathrm{H_2O}=10^3\ \mathrm{cm}^{-3}$
plotted over the same energy range.
The collision frequency increases with
energy in the oxygen-forming reaction, while the water-forming 
reaction is independent of energy and the OH-forming
reaction declines with energy.}
 \label{reacRates}
 \end{figure}

The cross sections (10$^{-16}$ cm$^2$) used in this paper to study reactions
\ref{h2o_chex}--\ref{o_chex} plotted in Fig.\ \ref{reacRates}a are given by
\begin{linenomath}
\begin{subequations}
\begin{align}
\mbox{Reaction \ref{h2o_chex}} :&\ \mathrm{H_2O + H_2O^+} \rightarrow
\mathrm{H_2O^+ + H_2O^*} \nonumber \\
 \sigma_\mathrm{H_2O} =&\ 38E_\mathrm{rel}^{-0.5} \label{h2o_chex_sigma} \\
\mbox{Reaction \ref{oh_chex}} :&\ \mathrm{H_2O + H_2O^+} \rightarrow
\mathrm{H_3O^+ + OH^*} \nonumber \\
 \sigma_\mathrm{OH} =&\   
38E_\mathrm{rel}^{-0.88}-0.39
\exp\left[{-\frac{1}{2}\left(\frac{E_\mathrm{rel}-57}{12}\right)^2}\right]
\label{oh_chex_sigma} \\
\mbox{Reaction \ref{o_chex}} :&\ \mathrm{H_2O + O^+} \rightarrow
\mathrm{H_2O^+ + O^*} \nonumber \\
 \sigma_\mathrm{O} =&\  
69E_\mathrm{rel}^{-0.29}+30
\exp\left[{-\frac{1}{2}\left(\frac{E_\mathrm{rel}-65}{18}\right)^2}\right].
 \label{o_chex_sigma}
\end{align}
\end{subequations}
\end{linenomath}
The Gaussian terms in Eqs.\
\ref{oh_chex_sigma} and \ref{o_chex_sigma} account for 
downward and upward trends in the associated data sets near 30 \kms,
but have little consequence on the neutral cloud,
given that most bound particles are produced at 
lower velocities.

Symmetric exchanges occur between like species by definition, although
unlike species also exhibit symmetric behavior on occasion.
Therefore, we explore several hypothetical behaviors for the OH$^*$-producing reaction
\ref{oh_chex} at low energies.
This test is separate from, but related to, the comparison between
reactions \ref{h2o_chex}--\ref{o_chex} themselves, and it motivates the point
that both high and low energy behaviors have an important effect on the neutral
cloud.
With \extrap, we have extropolated the best-fit curve (Eq.\ \ref{oh_chex_sigma}) 
to the lowest energies.
Symmetric and asymmetric behaviors are explored with \sym\ and \asym\ \citep{rapp1962,johnson1990}.
\sym\ is the same as \extrap\ except that below 1.5 eV,
$\sigma^\mathrm{symmetric}_\mathrm{OH}=30E_\mathrm{rel}^{-0.5}\times10^{-16}$ cm$^2$.
Notice that a similar energy dependence also applies to 
Eq.\ \ref{h2o_chex_sigma}, consistent with symmetric charge exchange.
\asym\ is the same as \extrap\ except that below 1.5 eV,
$\sigma^\mathrm{asymmetric}_\mathrm{OH}=11E_\mathrm{rel}^2\times10^{-16}$ cm$^2$.
Although it could be argued that $\sigma_\mathrm{OH}^\mathrm{symmetric}$ better fits 
the data if the two measurements at 2 eV are ignored,
our results for reaction \ref{oh_chex} were obtained with \extrap\ unless
noted otherwise.
We will discuss the implications of choosing
\extrap\ over \sym\ and \asym\ shortly.

The collision frequencies ($n\sigma v$) are plotted in Fig.\
\ref{reacRates}b for a given neutral density---in this case for $n_\mathrm{H_2O}=10^3$ cm$^{-3}$.  
The collision frequency for oxygen increases with relative speed, while
it is constant for water, and peaks at low velocities for OH.
The significance is that the oxygen
cloud tends to be more extended than either the OH or H$_2$O clouds.
The average collision frequency is also much higher for oxygen
($\times$10) than for either OH or H$_2$O, resulting in greater oxygen
abundance.
 
\begin{figure}
 \noindent\includegraphics[height=6.5in]{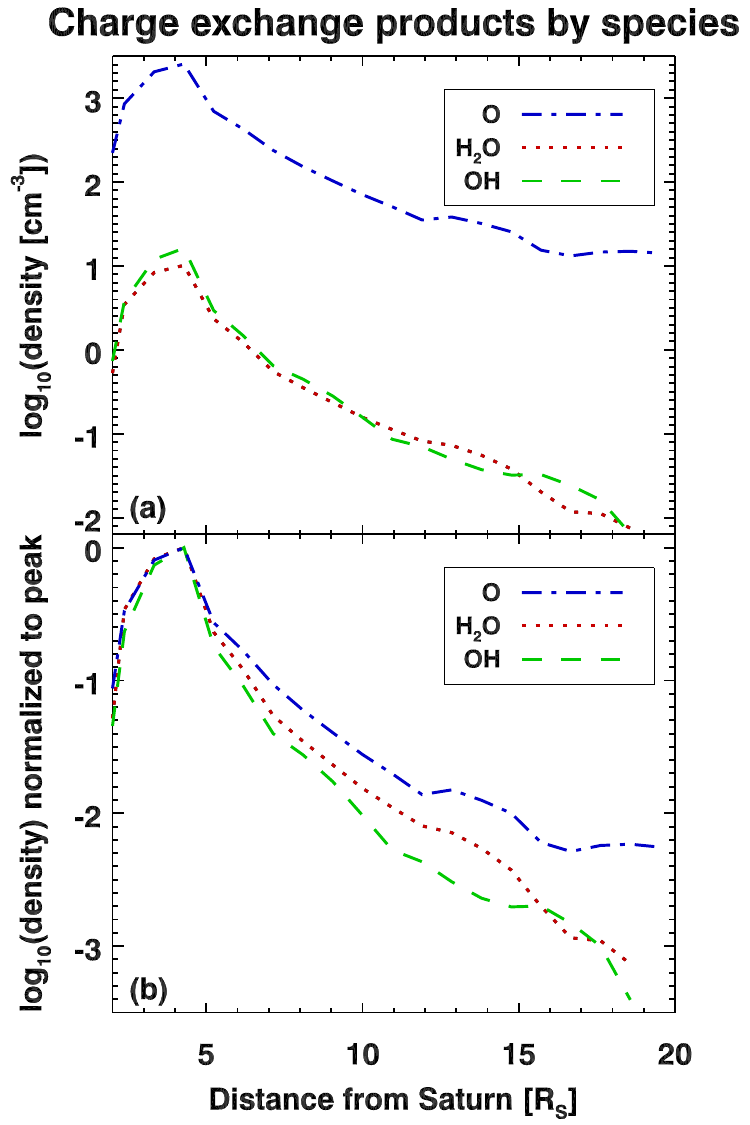}
 \caption{Neutral clouds produced by the reactions shown in Fig.\ \ref{reacRates}.
(a.)  Oxygen is the most abundant because the cross section section is
10$\times$ higher than than with O and OH.  The lifetime of oxygen
against photoionization
is also much longer than the lifetime for either OH or H$_2$O against
photodissociation.  (b.)  Same as above, but normalized to peak.  Oxygen
shows the most spreading because reactants are produced with higher velocities
(Fig.\ \ref{reacRates}b), which expands the cloud.  The same trend
holds with H$_2$O and OH, where OH tends
to be created with the lowest velocities (Fig.\ \ref{reacRates}, \extrap).} \label{chexCloud}
 \end{figure}

The equatorial neutral cloud densities resulting 
from reactions \ref{h2o_chex}--\ref{o_chex} are plotted in 
Fig.\ \ref{chexCloud}.  Only neutrals produced from charge exchange are shown;
neither the Enceladus neutral torus, nor the neutrals produced \textit{via} dissociation
have been included.  Oxygen is two orders of magnitude more
abundant than either OH or H$_2$O because of the higher rate of production,
but also because oxygen has a longer lifetime against
photoionization than either OH or H$_2$O have against photodissociation.
Unlike \cite{cassidy2010a}, dissociated neutrals from the latter processes are not tracked in our model.
Beyond the scope of the present study, this additional heating source would
serve to further inflate the oxygen and OH clouds.
Fig.\ \ref{chexCloud}b is the same as \ref{chexCloud}a, except
that the profiles are normalized to the peak density at the orbit of Enceladus.
The oxygen cloud is seen to be the most extended, followed by water, and finally
by OH,
with an order of magnitude separating the three species at 20\,$\mathrm{R_S}$.  

\begin{figure}
 \noindent\includegraphics[height=6.5in]{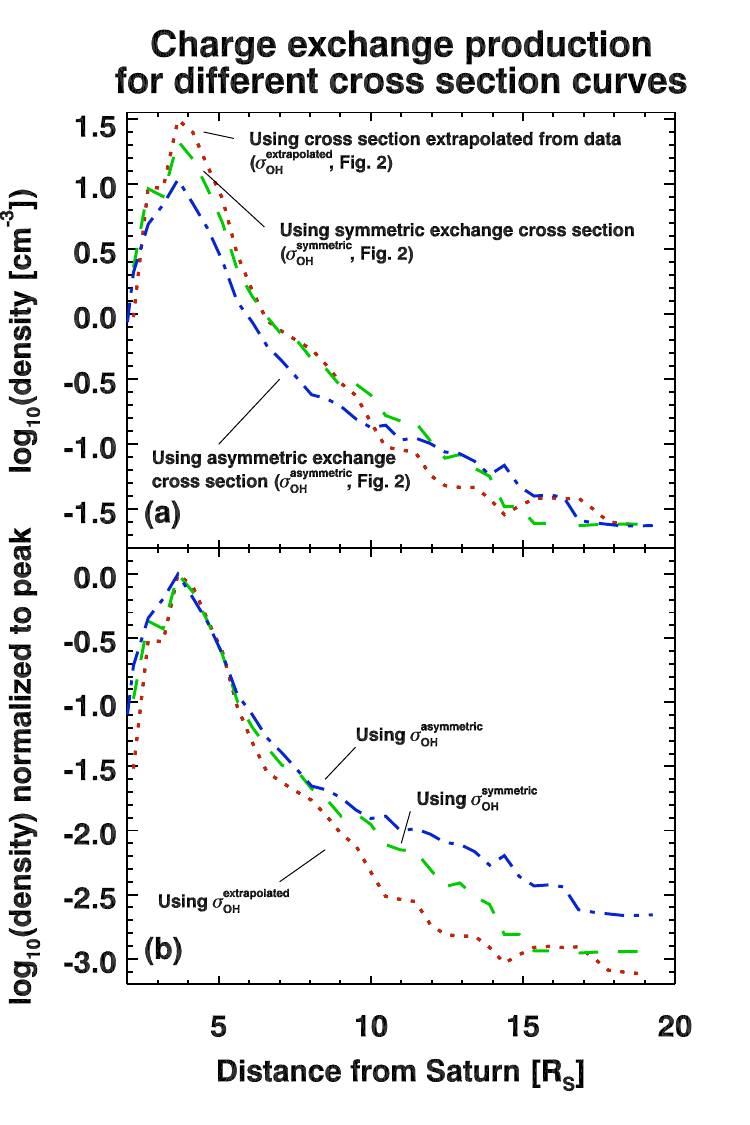}
 \caption{Neutral OH clouds produced from three hypothetical charge exchange cross
sections: \extrap, \sym, and
\asym\ (Fig.\ \ref{reacRates}).  (a.)  \extrap\ produces the highest density
(\asym, the lowest) at Enceladus because of the creation of additional
low-velocity particles.
(b.)  Same as above, but normalized to peak.  The
differences in density in the tail is not an indication of spreading, but rather
further illustrates the deficiency in the peak density, going from \extrap\ to
\asym.}
 \label{lowVelChexCloud}
 \end{figure}

The effects of low-velocity charge exchange are shown in 
Fig.\ \ref{lowVelChexCloud}.  
In Fig.\ \ref{lowVelChexCloud}a, 
we see that the peak density (as well as the total neutral cloud 
content) is the highest with \extrap\ because more low-velocity neutrals are produced 
than with either \sym\ or \asym. 
Conversely, fewer low-velocity neutrals are
available to populate the region near Enceladus's orbit with \asym\ 
when compared to either \extrap\ or \sym.
Stated another way, \extrap\ yields 
a neutral cloud with the steepest slope, and \asym, the shallowest.
Fig.\ \ref{lowVelChexCloud}b is identical to Fig.\ \ref{lowVelChexCloud}a, 
apart from normalization.  In this case, the slope of the density profile should
not be confused with the effect of inflating (spreading) the OH cloud.  It
should be viewed, rather, as the enhancement or depletion of low velocity
neutrals to fill the region inside of $\approx 10$ \Rs.  In other words,
neutrals beyond 10 \Rs\ are mostly formed in charge exchanges at high
velocities, for which all $\sigma_\mathrm{OH}$ converge to the same curve 
(Fig.\ \ref{reacRates}).
 
We have assumed to this point that the plasma is sub-corotating
in Enceladus's orbit (18 \kms, \cite{wilson2009}).  One might
expect, however, that the neutral clould would be affected in a measurable way
if instead, the plasma corotates at 26 \kms.  The H$_2$O cloud would
be least affected, given that the collision frequency of reaction \ref{h2o_chex} is independent
of speed (Fig.\ \ref{reacRates}b), but what about reactions such as \ref{oh_chex} and \ref{o_chex}, whose
collision frequencies are velocity-dependent?
Increasing the plasma speed amounts to shifting the spread of
ion velocities in Fig.\ \ref{reacRates} to the right, which would on average increase the speed
of the neutral products.
This is indeed the case, and in such a test
where we increased the plasma speed from 18 to 26 \kms, the oxygen cloud
increased in abundance and became even more extended.
The OH cloud also expanded somewhat, but decreased in total abundance.  Unfortunately, the
differences were less than 10\% in both the slope of the distribution and in
total oxygen abundance,
suggesting that neutral cloud observations are in this way unlikely to predict plasma
speeds in the torus.

\subsubsection{Neutral cloud sources:  plume $vs.$\ neutral torus}
We described in section \ref{neutralTorusProduction} the production in our model
of the neutral H$_2$O torus from the Enceladus plumes.  The plumes themselves
have also been prescribed as a separate
background density ($n_\mathrm{plume}$, section \ref{plume}) so
that we can
compare charge exchange occurring throughout the neutral torus to that occurring
only within the Enceladus plumes.

\begin{figure}
 \noindent\includegraphics[height=6.5in]{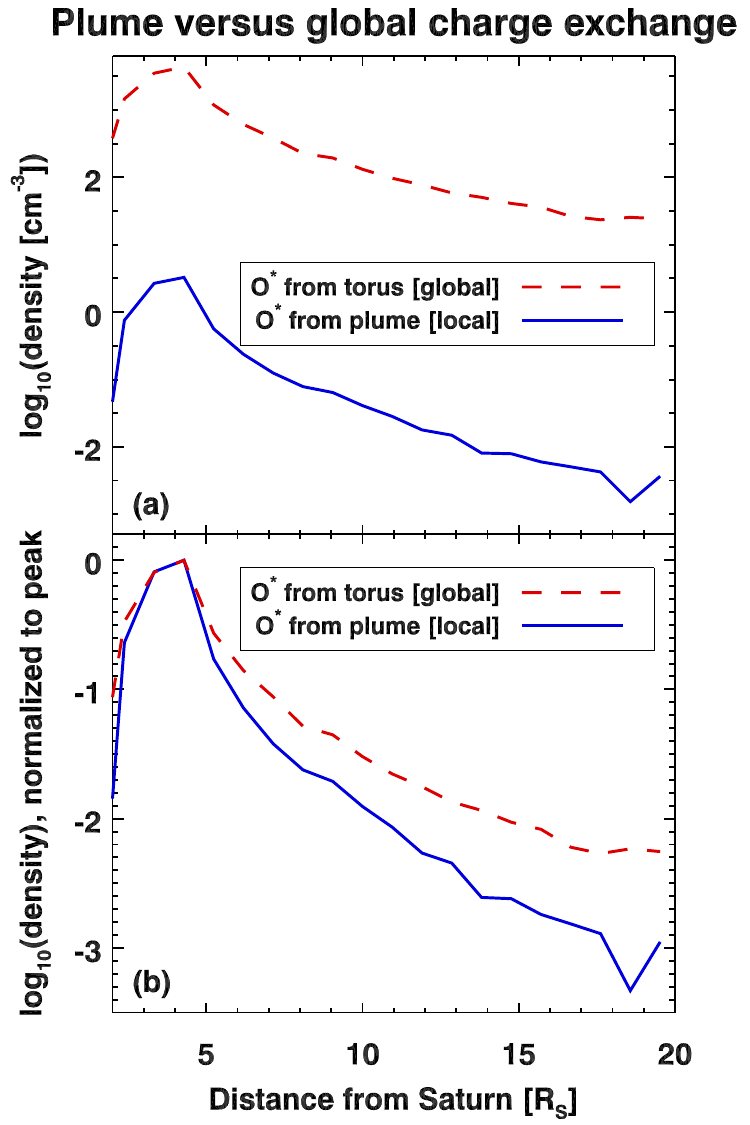}
 \caption{(a) Comparison between charge exchanged neutrals produced near the Enceladus plume 
and those produced from the neutral torus as a whole---in this case for oxygen.
(b.)  Though shown here for oxygen, all charge
exchange reactions near the plume result in a cloud with less spreading than
their global counterpart due to the imposed slowing of the plasma (and hence, the
release of slower neutral products) near the plume in
response to mass-loading \citep{fleshman2010a}.}
 \label{localVglobal}
 \end{figure}

The results are shown in
Fig.\ \ref{localVglobal}, where we have plotted the oxygen clouds produced
from charge exchange within both the Enceladus plumes (local) and the
entire neutral torus
(global).  The results are for reaction \ref{o_chex},
but the same test with reactions \ref{h2o_chex} and \ref{oh_chex} produces similar
results.
Immediately noticable is that the local production is $\approx0.1$\% of the overall
neutral production.  The torus's
dominance of neutral production can be explained as follows.  First, the volume
of the torus where reactions are occurring can be estimated as
$2\pi(4\mathrm{R_S})(0.2\mathrm{R_S})^2$, where
$0.1\mathrm{R_S}$ is roughly the torus's scale height.
The volume of the plume can be estimated from Eq.\ \ref{plume_density}, where
the dimensions are on the order of a cylinder with width 2$\mathrm{R_E}$ and
height $H_r\approx 16\mathrm{R_E}$.  Dividing these volumes gives roughly
$250(\mathrm{R_E}/\mathrm{R_S})^3\approx 10^{-5}$.
Further, the collision frequencies are proportional to the
neutral density, which in the plume are on the order of
10$^7$ cm$^{-3}$, whereas typical torus densities are 10$^5$ cm$^{-3}$, making
collisions in the plume 100$\times$ more frequent per volume than in the torus.
All told, the ratio of the volumes ($10^{-5}$) combined with the ratio of
densities ($10^2$) explain the local-to-global neutral production ratio of
$10^{-3}$ shown in Fig.\ \ref{localVglobal}a.
A similar pattern has been shown to exist at
Jupiter by \cite{bagenal1997} and \cite{dols2008}, where the majority of plasma is produced throughout 
the neutral torus, rather than near the interaction at Io itself.

The slopes of the neutral clouds from the plume and torus are
most easily compared in Fig.\ \ref{localVglobal}b, in which the
density profiles have been normalized.  The local source produces a more confined
neutral cloud because the ions from which they originate have been slowed near the
plume to account for the
effect of mass-loading (\cite{fleshman2010a}).  Nevertheless, such a signature
would be difficult to untangle in the data since global exceeds local production so
overwhelmingly.
  
\subsection{Dissociation} \label{results:dissociation} 
\begin{figure}
 \noindent\includegraphics[height=6.5in]{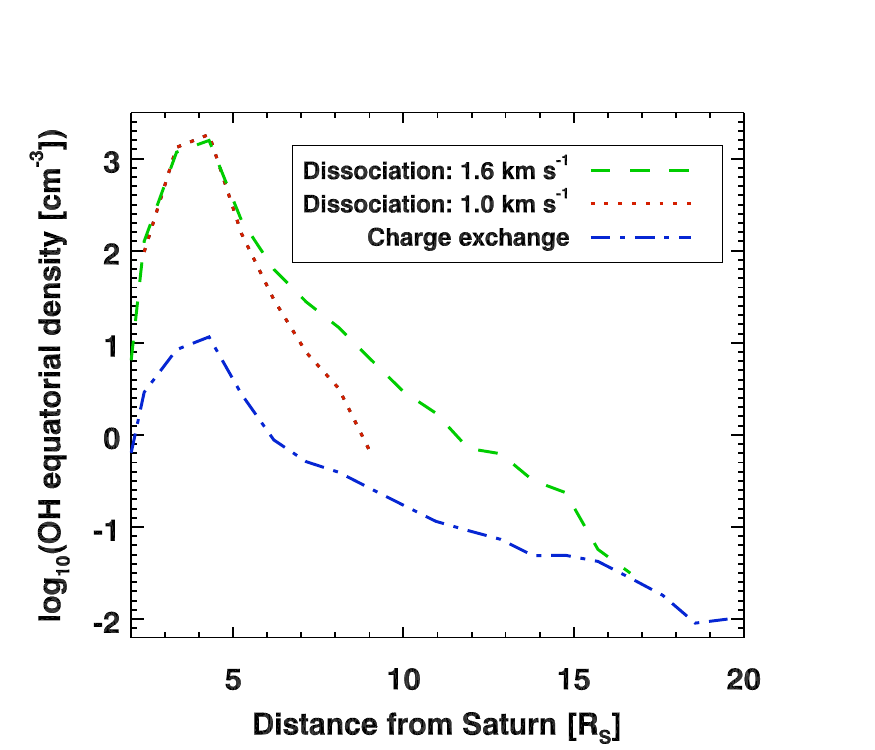}
 \caption{OH clouds produced from charge exchange 
and high- and low-speed 
dissociation.  Dissociation
dominates neutral cloud production inside 9--15 \Rs, at which point charge
exchange becomes the dominant contributor.}
 \label{ohCloud}
 \end{figure}

A major component of the OH cloud is produced by dissociation within the neutral torus,
whereby the initial velocities of the OH products range from 1 to 1.6 \kms\ \citep{wu1993,makarov2004}.
In Fig.\ \ref{ohCloud}, the clouds resulting from 
the high- and low-speed cases are plotted along with the
result from velocity-dependent charge exchange in section
\ref{chargeExchangeResults}.
First note that
dissociation contributes 100$\times$ more OH than does charge exchange 
at the orbit of Enceladus (4 \Rs);
the total cloud mass is almost 100$\times$ greater as
well.  Second, dissociation dominates 
over charge exchange from the Enceladus torus out to 9 and 15 \Rs\ in the low-
and high-speed cases, respectively.
The OH cloud content will only be marginally affected by variable solar activity
\citep{jackman2011}, given that impact dissociation contributes
4$\times$ more neutrals than does photodissociation, by virtue of the respective reaction rates
(section \ref{dissociation}).  In both cases, 
few neutrals are
absorbed by the rings, and even less by Saturn itself.  The same is not
true of charge exchange, where $\approx50$\% of the neutrals are
absorbed by Saturn (section \ref{results:neutralFates}).  

\begin{figure}
 \noindent\includegraphics[width=6.5in]{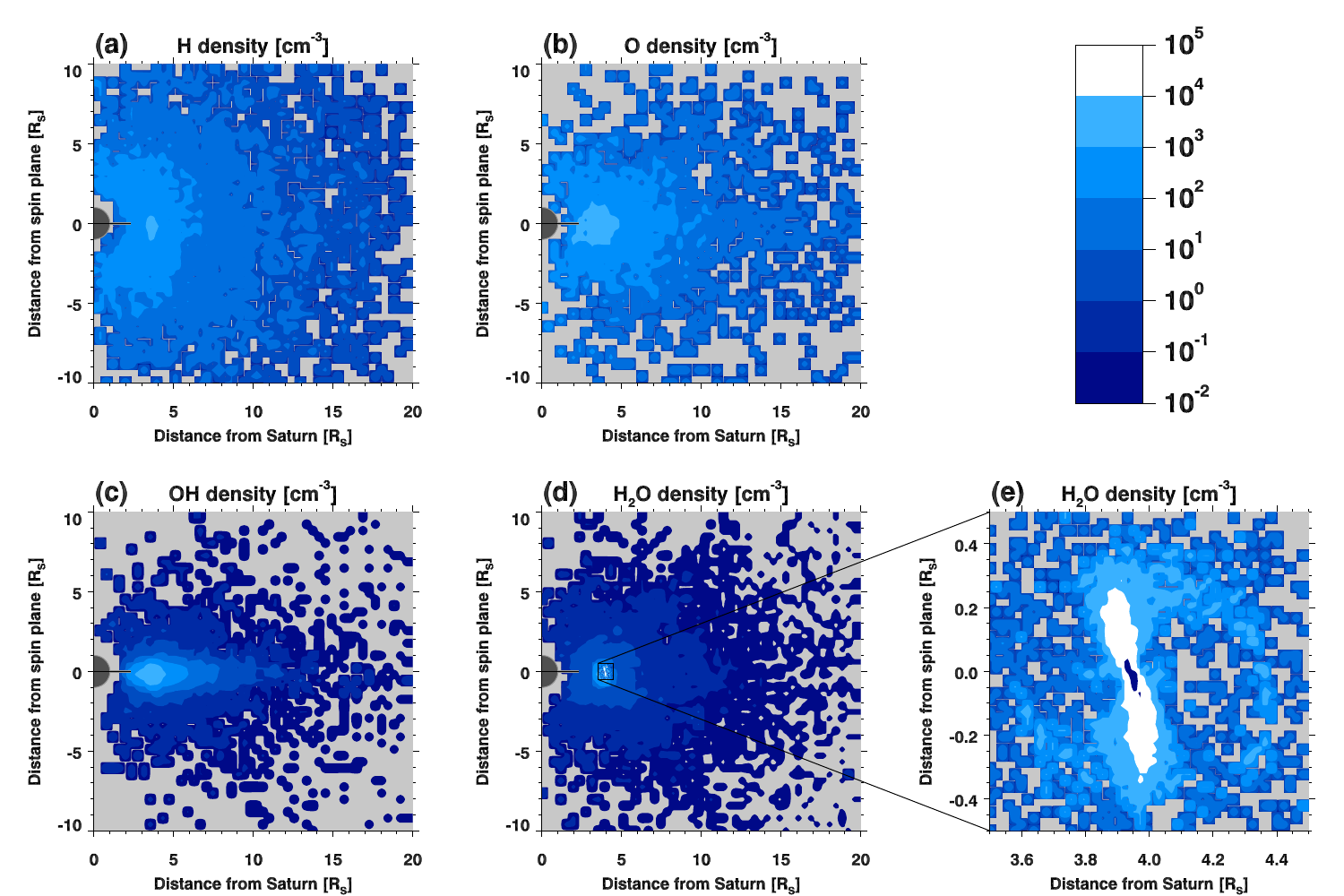}
 \caption{Neutral cloud densities in the $r$--$z$ plane.
(a.) Hydrogen produced purely from H$_2$O dissociation.
(b.) Oxygen produced purely from charge exchange (reaction
\ref{o_chex}).  
(c.)  Hydroxyl produced from the combination of charge exchange and
dissociation.  Dissociation dominates inward of 9--15 \Rs\ along the
equator, while charge exchange (reaction \ref{oh_chex})
tenuously fills the magnetosphere elsewhere.
(d.)  Water produced entirely by charge exchange (reaction \ref{h2o_chex}).
(e.)  Dense torus fed directly by the Enceladus plumes (section
\ref{plume}).}
 \label{allClouds}
 \end{figure}

Fig.\ \ref{allClouds}c is a two-dimensional version of Fig.\ \ref{ohCloud},
where the dissociation results have been
averaged and added to the results from charge exchange.
Saturn is at the left, and the Enceladus's orbit is located on the equator at 4 \Rs.
In addition to being confined radially, the dissociated neutrals
are also bound tightly to the equator, while
neutrals from charge exchange tenuously fill the magnetosphere.

Fig.\ \ref{allClouds}a shows the hydrogen cloud that accompanies the
dissociated OH clouds ($\mathrm{H_2O + e,\gamma}\rightarrow\mathrm{OH^*+H^*}$).
To conserve momentum, the hydrogen atoms have 17$\times$ the speed of the dissociated 
OH molecules, and thus
range between 17 and 27 \kms, with a relatively large, diffuse neutral cloud.
Shown is the result for the low-speed case, which produces more bound particles
and thus a more substantial neutral cloud.
Charge exchange from reactions such
as $\mathrm{H_2O+H^+}\rightarrow\mathrm{H_2O^++H^*}$ are also responsible for
H-cloud production, and deserve attention in future studies.

\subsection{Fates of neutral atoms and molecules} \label{results:neutralFates}
In our model, neutrals created by dissociation and charge exchange are
eventually either absorbed by Saturn, escape the system, or orbit until
they are destroyed (ionized) by photons.
In Fig.\ \ref{neutFates}a the fates for each
species are given by percentage.  In the case of hydrogen, the results
are from the dissociation model, described in section
\ref{results:dissociation}.  The enormous amount of escape (84\%) is due to the high
velocities ($\approx 17$ \kms) with which hydrogen is created following H$_2$O dissociation,
and the 8\% absorption is largely comprised of hydrogen
which would escape the system otherwise.

Oxygen is produced purely from charge exchange in our model (reaction \ref{o_chex}).  About
one-half escapes, one-third is absorbed, and the remaining 13\% 
contributes to the neutral cloud before being photoionized.  Water 
is also produced purely by charge exchange (reaction \ref{h2o_chex}) with 18\%
contributing to the neutral cloud.  Percentage-wise, more water is absorbed than oxygen
because oxygen is produced with higher speeds and generally larger orbits (section
\ref{chargeExchangeResults}).

The fate of OH is dominated by dissociation:
96\% feed the neutral cloud (ultimately ionized), 4\% are absorbed, and virtually none escape.  The reason
for the large percentage of bound and unabsorbed neutrals is that dissociated
OH has a velocity spread of 1 to 1.6 \kms\ in the neutral frame, compared to the 
escape speed of $\approx 5$ \kms\ in the same frame.
Looking only at OH produced by charge exchange (minor compared to dissociation),
58\% are absorbed, 23\% supply the neutral cloud, and 20\%
escape.  Compared to H$_2$O, an even greater percentage of charge-exchanged OH is absorbed 
because the cross sections favor production of
low-velocity OH molecules (Fig.\ \ref{reacRates}b).

The production of oxygen \textit{via} dissociation of H$_2$O has been ignored in
this paper on the grounds that, unlike OH, oxygen is largely produced by charge
exchange.  The cross section for oxygen-producing charge exchange is an order of
magnitude higher than that for the OH-producing reaction near the plasma flow
speed of $v_\mathrm{plasma}=18$ \kms\ (Fig.\ \ref{reacRates}a), while the
photodissociation rates are an order of magnitude smaller \citep{huebner1979}.
We estimate that including oxygen produced from dissociation would increase the
total oxygen cloud content by less than 20\%.

Charge exchange and dissociation play a large role in creating
Saturn's neutral clouds from the plume-fed neutral torus. 
The reactions we have included have been chosen to demonstrate the effects of
low velocity charge exchange and dissociation,
but they are also among the most important.  The
neutral cloud densities presented in this paper are expected 
to undershoot the results from models which include the additional
reactions found in Fig.\ 3, of \cite{fleshman2010b}
by no more than a factor of two.
With this caveat in mind, we now
compare the present results with several other recent models.

\subsubsection{Comparison with other models} \label{comparisonWithModels}
\begin{figure}
 \noindent\includegraphics[width=6.5in]{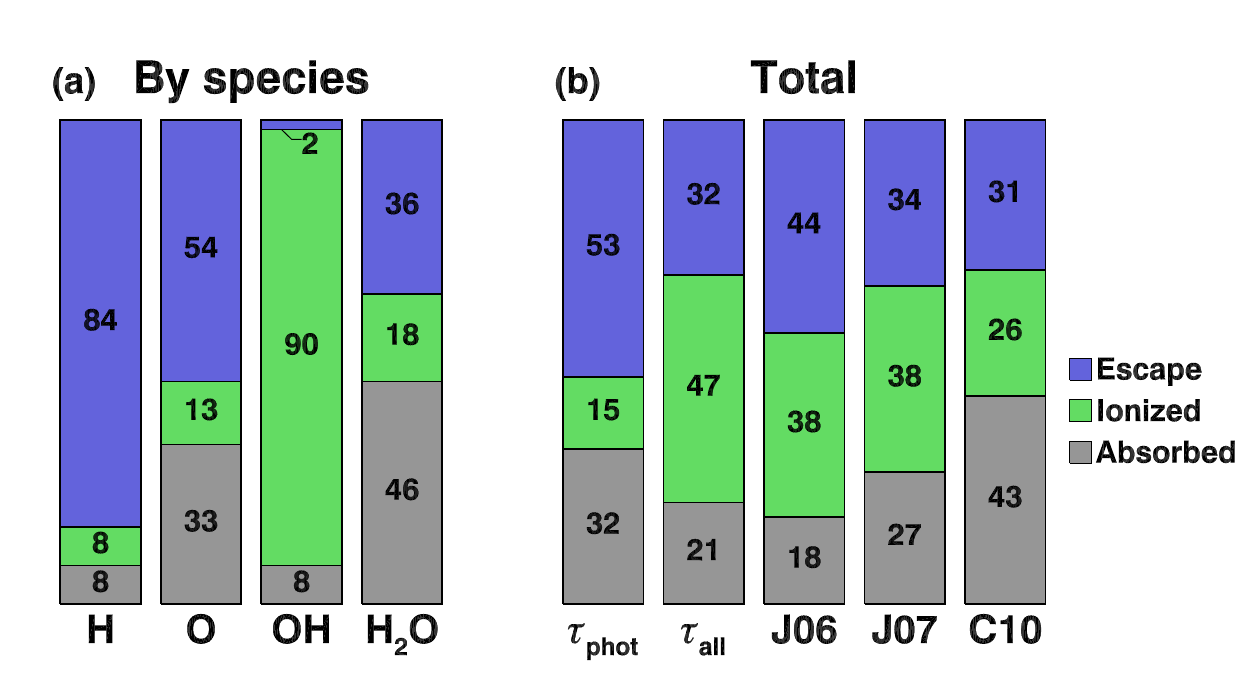}
 \caption{The fates of neutrals in our model along with the results from other
models.  (a.)  Dissociation produces low-velocity neutrals and
OH is thus not likely to escape or to be absorbed.
Conversely, dissociation also produces hydrogen which largely leaves the
system.  (b.)  The results of 
J06 \citep{johnson2006}, J07 \citep{jurac2007}, and C10 
\citep{cassidy2010a}, along with our own weighted totals (excluding hydrogen;
see section
\ref{comparisonWithModels}).
In the case of $\tau_\mathrm{phot}$, the lifetime
of the cloud is determined by photoionization/dissociation only, whereas
with $\tau_\mathrm{all}$, we
limit the lifetimes by also including electron impact and charge exchange.
These limiting cases bound the previous studies, except
that C10 has more absorption attributed to neutral--neutral collisions.}
 \label{neutFates}
 \end{figure}

Fig.\ \ref{neutFates}b:\ J06 is the work of \cite{johnson2006}, where they also investigated 
the neutral clouds created from
low-velocity charge exchange in the stagnated flows in Enceladus's orbit.
Fig.\ \ref{neutFates}b:\ J07 is from \cite{jurac2007}, where the authors were primarily interested in the
interaction between the neutral cloud and Saturn's rings.  The most recent
model comes from \cite{cassidy2010a} (C10), where they investigated the
spreading of the neutral cloud from neutral--neutral
collisions.

To compare with these studies, we first had to weight our H, O, OH, and H$_2$O clouds.
We did this for two limiting cases.
In the first case ($\tau_\mathrm{phot}$, Fig.\ \ref{neutFates}b),  we assume, as we have thus far, 
that the neutral clouds evolve until destroyed by
either photoionization or photodissociation: H, O, OH, H$_2$O =
 40, 14, 0.6, 0.3 years, respectively.  These lifetimes yield an upper limit
since charge exchange and electron
impact are not included as losses.  In the second case
($\tau_\mathrm{all}$) , we derived a lower limit to the 
lifetimes from Table 2
of \cite{fleshman2010b} by
summing the additional losses due to 
charge exchange and electron impact, finding:
H, O, OH, H$_2$O = 0.4, 0.4, 0.2, 0.03 years, respectively.
Notice in particular the drastically different times scales for H and O,
where including the additional sinks reduce the size of the
H cloud by a factor of 40/0.4 = 100, and the oxygen cloud by 14/0.4 = 35.  This
case represents an extreme limit, given that the neutrals spend almost all of their
time orbiting outside of the Enceladus torus, where compared to photo-processes,
the chances of charge exchange and electron
impact are relatively unlikely.  
We mention, however, that \cite{rymer2007,rymer2008} has shown that circulation patterns
inside of 12 \Rs\ at Saturn gives rise to `butterfly' hot electron pitch angle
distributions, related to low temperature anisotropy ($T_\perp/T_\parallel$), on
which proton field-aligned distributions depend \citep{sittler2008}.

The individual clouds (excluding hydrogen) were weighted by the stated
time scales and totaled
in Fig.\ \ref{neutFates}b.  When only losses to photodissociation/ionization are
considered ($\tau_\mathrm{phot}$), the neutral cloud is dominated by oxygen,
whose fate thus determines that of the neutral cloud.  When 
charge exchange and electron impact are also included ($\tau_\mathrm{all}$),
dissociated OH contributes significantly, driving 
the neutral cloud (ionized) percentage up, and the escape percentage down.
We note that the neutral fates presented in \cite{bagenal2011}
(escape = 44\%, ionized = 17\%, absorbed = 39\%) were
based on an earlier version of our model which only included H$_2$O.

The particles that are neither absorbed nor lost by escape make up the neutral
clouds.
In the case where the cloud evolves for $\tau_\mathrm{phot}$, oxygen
and hydrogen dominate since they are far less likely to be photoionized than are
OH and H$_2$O to be photodissociated.  With charge
exchange and electron impact included ($\tau_\mathrm{all}$), however, more oxygen
and hydrogen are removed from the system, which then tends to favor a molecular OH--H$_2$O
cloud.  In terms of total mass the same applies, although hydrogen
accounts for only a few percent at most.  We find that the total cloud mass is
bounded between $\approx1$ and 10 Mtons, for $\tau_\mathrm{all}$ and
$\tau_\mathrm{phot}$, respectively.

It is worth pausing to re-emphasize that the system is in reality better represented  by the
$\tau_\mathrm{phot}$ case, from which all neutral clouds in this paper have been
derived.  The $\tau_\mathrm{all}$ case is strictly valid only for neutrals
within the Enceladus torus, though reactions with electrons and protons may also
prove important, as discussed above.  What is illustrated, however, is
that Saturn's magnetosphere is less 
oxygen-dominated than suggested by looking at losses from photo-processes
alone.  These results suggest
that our oxygen abundances are somewhat overestimated, likely by less than a factor of
two.

\subsubsection{Neutral absorption}
\begin{figure}
 \noindent\includegraphics[width=6.5in]{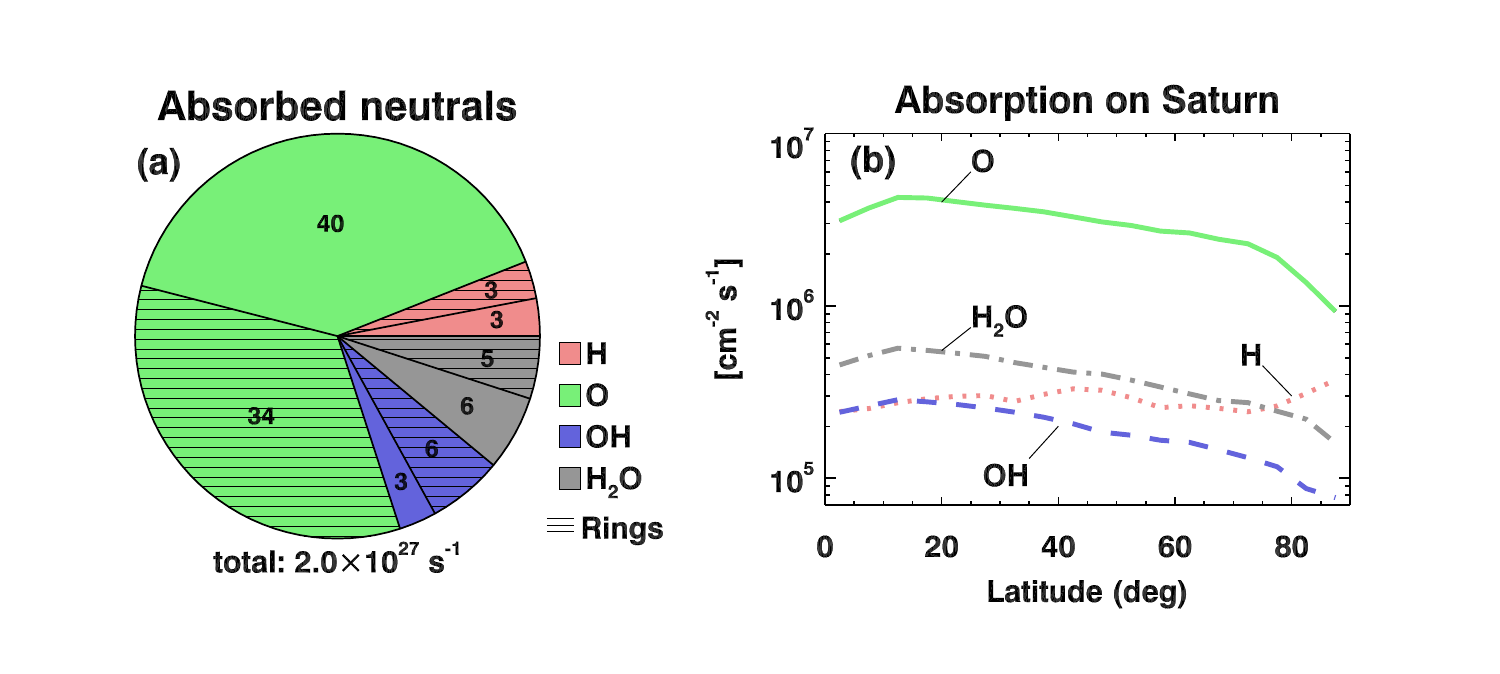}
 \caption{(a.)  Neutrals absorbed by Saturn, plotted by species.
Partitions with horizontal lines
indicate percentages absorbed by Saturn's rings.  (b.)  Neutral flux
on Saturn as a function of latitude.  Neutrals produced by charge exchange
(H$_2$O, OH, and O) peak in flux at low latitudes due to the nature of the ion
distributions from which they orginate, which have initial velocity vectors
predominantly in the ring plane.  Conversely, hydrogen flux is constant
across Saturn because it originates from dissociation, whose velocity
distribution is prescribed as isotropic.
Note that OH produced by dissociation is not energetic
enough to reach Saturn.}
 \label{absorptionFates}
 \end{figure}

The particles absorbed by Saturn and its rings are plotted by species and
latitude in Fig.\ \ref{absorptionFates}.  In Fig.\ \ref{absorptionFates}b, we see that most
absorption comes from oxygen (74\%), followed by H$_2$O (11\%), OH (9\%),
and finally by hydrogen (6\%).  Absorption is equally divided
between Saturn and its rings except in the case of OH, where twice as much falls
on Saturn's rings.  This is because OH is largely produced by impact
dissociation, which creates slower neutrals than does charge exchange, whereby
in our model, H$_2$O and oxygen arise exclusively.

In Fig.\ \ref{absorptionFates}b, absorption is plotted
against Saturn's latitude.  Because the model is symmetric about the equator,
the results apply to either hemisphere.  Oxygen, water, and OH follow the same
trends because they all originate from charge exchange (dissociated OH is slow
and does not
reach Saturn), and have
been created from ions with similar velocity distributions.  Any second-order
differences due to the velocity-dependence of the respective cross sections are not
immediately apparent.  Hydrogen, on the other hand, is produced entirely by dissociation in
the model and exhibits a more uniform flux across Saturn.  The explanation is that 
the velocity distribution from which hydrogen is produced is isotropic, whereas
that which produces charge-exchanged neutrals is bi-Maxwellian 
(section \ref{chargeExchange}).
The fluxes shown in Fig.\ \ref{absorptionFates}b are consistent with \cite{hartogh2011}, 
who modeled recent Herschel observations of
Saturn's water torus and found an average flux of 
$6\times10^5$ cm$^{-2}$ s$^{-1}$ for H$_2$O + OH impinging on Saturn.

\section{Discussion} \label{discussion}
Some useful conclusions can be drawn by further contrasting our results with
\cite{cassidy2010a} (C10).
It is important that we first mention a profound difference between our models.
The model of C10
effectively carries out resonant charge exchange only, which does not chemically
alter the neutral population; neutrals in their model are produced either directly from
Enceladus or from subsequent dissociations.  Neutrals in our model, on the other
hand, originate from Enceladus (H$_2$O).  OH is then created $via$
dissociation (as with C10), but secondary O, OH, and H$_2$O populations are
$created$ from H$_2$O $via$ charge exchange with the dense plume-fed Enceladus torus.  The
C10 model redistributes neutrals around Saturn, while we redistribute and
chemically re-assign
neutral abundances by allowing for asymmetric charge
exchanges.  Thus, it may well be a coincidence that our models are similar
in total abundance.  While it may be difficult to compare our total abundances,
the slope of our radial density profiles can be contrasted
directly
because our redistribution mechanisms (charge exchange and
dissociation) are similar.  Differences are due largely to C10's inclusion of neutral
collisions
and our prescribing unique velocity-dependent charge exchange for each of the
O-, OH-,
and H$_2$O-producing reactions (reactions \ref{h2o_chex}--\ref{o_chex}).

\begin{figure}
 \noindent\includegraphics[width=4.5in]{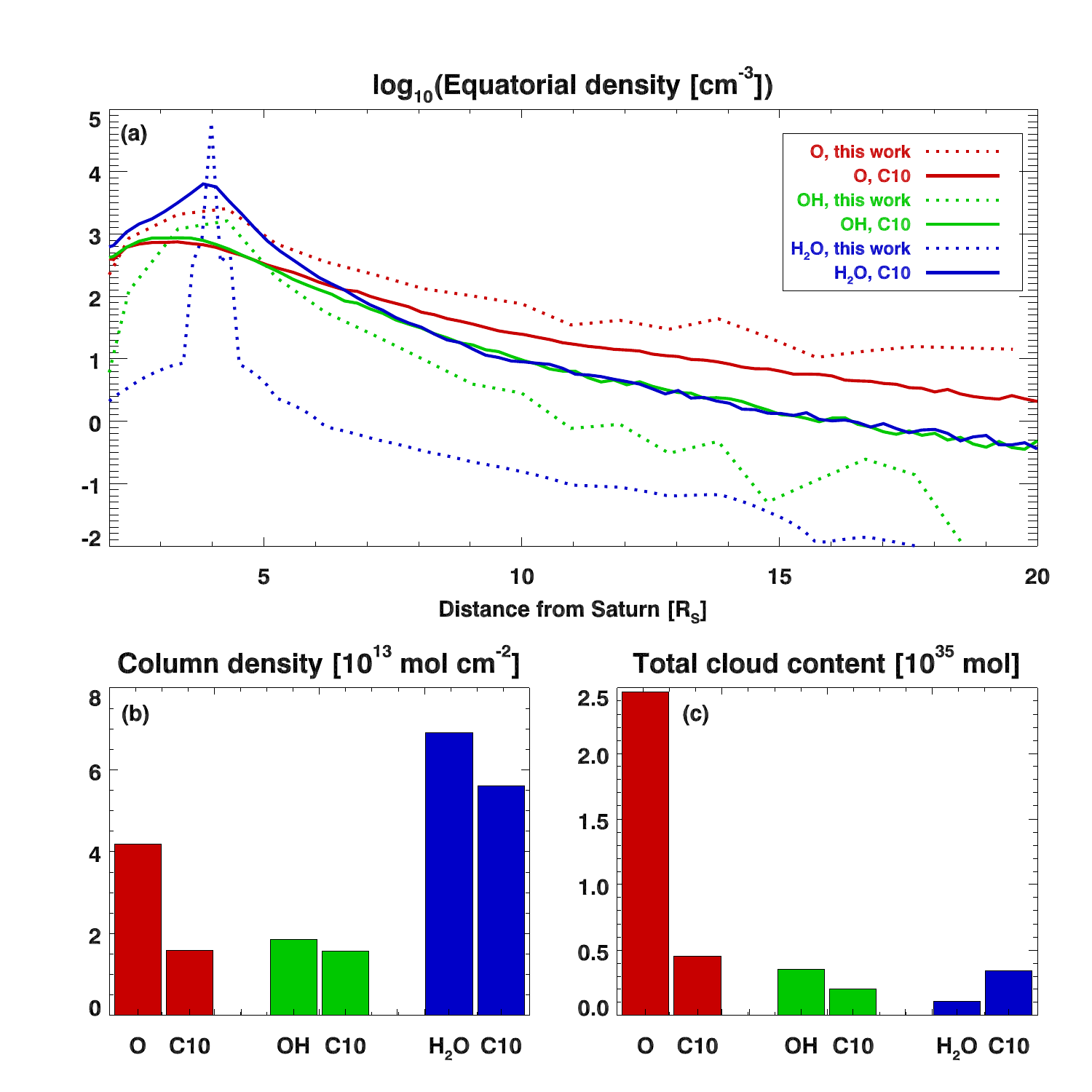}
 \caption{
(a.)  Total neutral clouds from our model, compared with \cite{cassidy2010a}
(C10).
All clouds include contributions from charge exchange (reactions
\ref{h2o_chex}--\ref{o_chex}), while H$_2$O is largely
comprised of water sourced directly from Enceladus, and OH includes
contributions from dissociation.
The cloud densities are limited by photodissociation for OH and
H$_2$O and by photoionization for O.
Including charge 
exchange as a loss for cloud neutrals would reduce the lifetime for O
more than for either the OH or H$_2$O, and would lower the relative oxygen
abundance accordingly.
(b.)  Equatorial column densities found by integrating the 
plotted equatorial densities.
The H$_2$O column density is similar to C10,
despite their having a very different
radial distributions.
(c.)  Total neutral cloud content.  Our total H$_2$O content
is less than C10 found, while our H$_2$O column density is higher
because our H$_2$O cloud is not subjected to neutral collisions and is thus more
confined.}
 \label{fleshmanVcassidy}
 \end{figure}

Our neutral clouds are compared with C10 in Fig.\
\ref{fleshmanVcassidy}.
All of our clouds include contributions
from charge exchange, but the H$_2$O cloud is mostly comprised of water sourced
directly from Enceladus (3.95\,\Rs), and OH includes the additional source from dissociation.
In the C10 model, the water molecules were spread due to neutral--neutral
collisions, which explains our higher H$_2$O 
densities near Enceladus's orbit (Fig.\ \ref{fleshmanVcassidy}a).
The slope of the oxygen profile agrees best with C10 because their charge
exchange cross section
most resembles our own (Eq.\ \ref{o_chex_sigma}).  Our H$_2$O profiles
agree less, and our OH slopes, the least, due mainly to the strong effect that
neutral collisions have on those more polar molecules.  In particular, C10 used
a much larger cross section for neutral collisions involving
H$_2$O and OH \citep{teske2005} than for atomic
oxygen \citep{bondi1964}.  This helps to further explain our agreement 
with their oxygen profile since we exclude neutral--neutral
collisions from our model altogether.
We conclude that neutral--neutral collisions appear to play a
less significant role with atomic species, such as
oxygen and hydrogen.

The column densities (Fig.\ \ref{fleshmanVcassidy}b) 
are similar to C10, who constrained their
O and OH clouds with the most recent Cassini UVIS results of \cite{melin2009}.
Our oxygen density---as well as our total oxygen content (Fig.\
\ref{fleshmanVcassidy}c)---is higher for two reasons.  First, we use 
a larger cross section than does C10
for reaction \ref{o_chex}, and second, the clouds presented here have been
limited only by photoionization.  Charge exchange and electron impact
are second order losses beyond 6 $\mathrm{R_S}$, but including them would 
favorably reduce the oxygen content more than 
OH and H$_2$O (section \ref{results:neutralFates}), bringing our models
into better agreement.

Our total H$_2$O content is 4$\times$ less than C10 found (Fig.\
\ref{fleshmanVcassidy}c).  This is partly because we have subjected 
H$_2$O molecules in the primary (plume-fed) neutral torus to the shortest lifetimes possible (section
\ref{enceladusTorus}), whereas C10 tracks molecules
that get kicked out of the densest plasma $via$ neutral collisions, and thus
survive longer, being less susceptible to both charge exchange and electron
impact.  That their total H$_2$O content is higher than ours (Fig.\
\ref{fleshmanVcassidy}c),
 does not contradict the fact that their H$_2$O column density is lower;
neutral--neutral collisions would spread out the torus,
lowering the column density, while allowing neutrals to survive longer,
increasing the total abundance.

Our model would benefit by including the redistribution
attributed to neutral collisions by allowing particles to interact in a
direct simulation Monte Carlo (DSMC) model such as in C10.  Likewise, DSMC models
would benefit by including charge exchange cross sections specific 
to each reaction.
Such models should also take into account asymmetric charge exchanges, which
affects neutral cloud composition.

The reactions modeled in this study were chosen
in order to measure the effect of symmetric and asymmetric charge exchange at
low velocities.
Building upon our findings, future studies 
should include additional neutral-producing charge exchanges, such as 
$\mathrm{OH^+ + H_2O} \rightarrow \mathrm{OH^* + H_2O^+}$,
$\mathrm{H^+ + H_2O} \rightarrow \mathrm{H^* + H_2O^+}$,
and $\mathrm{OH^+ + H_2O} \rightarrow \mathrm{O^* + H_3O^+}$, as well as
dissociative recombination of H$_2$O$^+$.

\section{Conclusions} \label{conclusions}
We have modeled low-velocity charge
exchange from the point of view of the ions, allowing us to study
the effects of velocity as well as gyrophase.
With reactions \ref{h2o_chex}--\ref{o_chex}, we have been able to 
offer an estimate on the size and shape of the neutral clouds at Saturn, while
simultaneously exploring the sensitivity of the neutral clouds to
a variety of velocity-dependent reactions.  

We have also re-visited the production of OH following H$_2$O dissociation in
the primary neutral torus.
Previous models have used 1 \kms\ as the initial velocity for OH, while
measurements suggest a range of speeds from 1 to 1.6 \kms.  In our model, the
higher speed increases the range within which dissociation dominates neutral
production from 9 to 15 \Rs.

Additional findings are:

(1.)  Charge exchange cross sections that increase steeply at
low speeds tend to produce neutral clouds more confined to the orbit of
Enceladus, implying the most spreading for oxygen,
moderate spreading for H$_2$O,
and the least for OH (Fig.\ \ref{chexCloud}).  Accounting for gyrophase 
doubles the local OH density within Encelacus's orbit, has $\approx$ no effect on
H$_2$O, and decreases oxygen density by less than 10\%.

(2.)  Enceladus is solely responsible for the creation of the neutral H$_2$O
torus $via$ thermal ejection from its plumes.  However, Saturn's neutral clouds are
overwhelmingly produced by charge exchange and dissociation occurring throughout the
torus (99\%),
and not near Enceladus itself (Fig.\ \ref{localVglobal}).
 
(3.)  We estimate that roughly half of all neutrals escape the system, with
the remaining equally divided between absorption by the rings/planet and the
neutral clouds
(Fig.\ \ref{neutFates}).  Less than 50 \kgs\ is thus
ionized and transported out of the system as plasma.  
This number is expected to represent an upper limit,
given we have assumed that all particles forming the neutral clouds are
ultimately ionized; a more accurate result would require modeling
the detailed effects of
charge exchange and neutral--neutral collisions within the neutral
clouds.
This estimate
can be compared to \cite{sittler2008}, whose Figs.\ 14 and 17 give roughly
$([NL^2]_\mathrm{W^+}/L^2)\times m_\mathrm{W^+}/\tau_\mathrm{transport}\approx
3\times10^{31}\times m_\mathrm{W^+}/10^5$ s $\approx 10$ \kgs\ at $L=10$.

(4.) 
Saturn's neutral cloud has a total mass of at least 1 Mton, but likely much
closer to 10 Mtons.  The primary plume-fed neutral torus (0.3 Mtons) is comprised entirely
of water in our model, while the secondary neutral clouds are
broken down into H ($\lesssim5$\%), O ($\lesssim82$\%), OH ($\gtrsim13$\%), and H$_2$O
($\approx1$\%).
Atomic oxygen dominates the composition both because of a high production rate
from charge exchange as well as a long lifetime against photoionization.
Charge exchange
and reactions with electrons favorably remove hydrogen and oxygen,
but are secondary loss mechanisms 
throughout the majority of the magnetosphere.

(5.)  Our model predicts fluxes on Saturn from charge exchange of
 $\approx6\times10^5$
cm$^{-2}$ s$^{-1}$ for both OH and H$_2$O  (consistent with Herschel observations
by \cite{hartogh2011}), and oxygen is about 5$\times$ higher.  Absorption is divided
equally between Saturn and its rings (Fig.\ \ref{absorptionFates}a).

(6.)  Our total neutral abundances are similar to \cite{cassidy2010a} (C10) for both
OH and H$_2$O, and 4$\times$ higher for oxygen (Fig.\ \ref{fleshmanVcassidy}).
Differences in the slopes of our equatorial density profiles are in part due to our not
including neutral--neutral collisions, while this fact appears to have no effect
on the oxygen profile.  On the other hand, C10 did not include 
the effects on neutral chemistry following asymmetric
charge exchanges, nor did
they use velocity-dependent cross sections particular to each reaction.
Herschel observations by \cite{hartogh2011} confirm the importance of neutral--neutral
collisions for H$_2$O, but if oxygen is the dominant neutral
species in Saturn's magnetosphere, as our model predicts, neutral--neutral collisions may
play a smaller role in Saturn's neutral cloud than previously expected.  

Given the effect on both the size and shape of the neutral clouds, 
we suggest that future neutral cloud models 
include charge exchange cross sections unique to each reaction.
Asymmetric charge exchange also
has an important effect on neutral chemistry that should be implemented.
Regarding the ions' gyrophase, Monte Carlo models can account for its effect by 
using phase-dependent probability distributions.
Finally, the range of OH velocities studied here should be considered when
modeling dissociation.

Moving forward, we plan to implement these suggestions into the neutral cloud model of
C10 and to couple that model with the
chemistry model of \cite{fleshman2010b}.  Constrained by Cassini plasma
observations,
the chemistry model uses C10's neutrals as input, and provides
ion temperatures and densities throughout the magnetosphere ($<$ 20 \Rs), which 
C10 in turn uses to update neutral densities.
An improved understanding of two issues is planned:  
(1) Where does plasma transport become important? (2) What is the role of hot
electrons with regard to ion--neutral chemistry inside 20 \Rs?


%
%
%
\appendix
\section{Collision probability}
The average number of collisions occurring during a time interval $\Delta t$ is 
given by $\lambda\equiv\nu \Delta t$, where
$\nu=n_\mathrm{neutrals}\sigma(v_\mathrm{rel})v_\mathrm{rel}$ is the local
collision frequency, assumed to be constant during $\Delta t$.
Statistics are applied to
determine if and how many reactions occur during $\Delta t$.
The Poisson distribution \citep{zwillinger1996,reif1965} gives the probability 
of suffering exactly $n$ collisions for a given $\lambda$:
\begin{linenomath}
\begin{equation} \label{appendix:f}
  f(n;\lambda)=\frac{e^{-\lambda}\lambda^n}{n!}.
\end{equation}
\end{linenomath}
Notice that Eq.\ \ref{appendix:f} peaks at $n=\lambda$
if one treats $n$ as a continuous variable.
Summing Eq.\ \ref{appendix:f} discretely from $n=k$ to $n=\infty$
gives the probability of
suffering \textit{at least} $k$ collisions during $\Delta t$,
\begin{linenomath}
\begin{equation} \label{appendix:pn}
P_k(\lambda)=e^{-\lambda}\sum_{n=k}^\infty \frac{\lambda^n}{n!}.
\end{equation}
\end{linenomath}
Because Eq.\ \ref{appendix:f} is normalized 
($e^{-\lambda}\sum_{n=0}^\infty \lambda^n/n!=e^{-\lambda}e^\lambda=1$), 
Eq.\ \ref{appendix:pn} can be conveniently rewritten as
\begin{linenomath}
\begin{equation} \label{appendix:pn2}
P_k(\lambda)
=1-e^{-\lambda}\sum_{n=0}^{k-1} \frac{\lambda^n}{n!}.
\end{equation}
\end{linenomath}
A random number ($0<N<1$) is compared to each $P_k$ at each timestep.  The
largest $k$ for which $P_k > N$ determines how many fast neutrals (collisions), $k$, are
produced during $\Delta t$.  

In practice, it is only necessary to compare to the
first few $P_k$ when $\lambda \ll 1$, made evident by the leading terms in Eq.\ \ref{appendix:pn} for
$k+1$ and $k$:
\begin{linenomath}
\begin{align} \label{appendix:ratio}
\frac{P_{k+1}}{P_k} \approx
  \frac{f(k+1;\lambda)}{f(k;\lambda)}
            = \frac{\lambda^{k+1}/(k+1)!}{\lambda^k/k!}=\frac{\lambda}{k+1}
\xrightarrow[\lambda\rightarrow0]{}0.
\end{align}
\end{linenomath}
Multiple collisions are thus increasingly unlikely 
when $\lambda \ll 1$.  In such cases, comparison with
$P_1=1-e^{-\lambda}\approx \lambda = \nu \Delta t$ is sufficient.
%
%
%
%

%
%
%
%

\begin{acknowledgments}
This work was supported under the NESSF program, fellowship number
11-Planet11R-0005.  BF thanks two reviewers for their feedback and many useful suggestions.
\end{acknowledgments}

%
%
%
%
%
%
%
%
%
%




%




%
%

\end{article}

%



%
 \end{document}